\documentclass[aps, prd, amsmath, floats, floatfix, twocolumn, nofootinbib, superscriptaddress, showpacs]{revtex4-1}
\usepackage{graphicx,color}
\usepackage{latexsym}
\usepackage{amsmath,amsfonts,amssymb}
\usepackage{physics}     % convinient physics shortcuts

\def\bC{{\mathbb C}}

\begin{document}

% TITLE

\title{Superradiance in the BTZ black hole with Robin boundary conditions}

% AUTHORS

\author{Claudio Dappiaggi}
\affiliation{Dipartimento di Fisica, Universit\`a degli Studi di Pavia, Via Bassi, 6, 27100 Pavia, Italy}
\affiliation{Istituto Nazionale di Fisica Nucleare -- Sezione di Pavia, Via Bassi 6, 27100 Pavia, Italy}

\author{Hugo R. C. Ferreira}
\affiliation{Istituto Nazionale di Fisica Nucleare -- Sezione di Pavia, Via Bassi 6, 27100 Pavia, Italy}

\author{Carlos A. R. Herdeiro}
\affiliation{Departamento de F\'isica da Universidade de Aveiro and Center for Research and Development in Mathematics and Applications (CIDMA), Campus de Santiago, 3810-183 Aveiro, Portugal}

% DATE

\date{October 2017}

\begin{abstract}
We show the existence of superradiant modes of massive scalar fields propagating in BTZ black holes when certain Robin boundary conditions, which never include the commonly considered Dirichlet boundary conditions, are imposed at spatial infinity. These superradiant  modes are defined as those solutions whose energy flux across the horizon is towards the exterior region. Differently from rotating, asymptotically flat black holes, we obtain that \emph{not} all modes which grow up exponentially in time are superradiant; for some of these, the growth is sourced by a bulk instability of AdS$_3$, triggered by the scalar field with Robin boundary conditions, rather than by energy extraction from the BTZ black hole. Thus, this setup provides an example wherein Bosonic modes with low frequency are pumping energy into, rather than extracting energy from, a rotating black hole. 
\vspace*{10ex}
\end{abstract}

\maketitle

% INTRODUCTION

\section{Introduction}

The beautiful phenomenon of superradiance found a generic realisation in black hole (BH) physics: the ability of extracting energy and angular momentum from rotating BHs --- see Ref.~\cite{Brito:2015oca} for a recent review and Ref.~\cite{Finster:2007fd} for a mathematically rigorous treatment. This extraction is mediated by low frequency Bosonic modes, which get amplified upon scattering off the rotating BH and grow exponentially, becoming an instability if trapped in the vicinity of the BH.  Much work has been devoted in the last decades towards understanding this instability, when the trapping is enforced by a mass term of the Bosonic field~(see $e.g.$~\cite{Damour:1976kh,Zouros:1979iw,Detweiler:1980uk,Furuhashi:2004jk,Dolan:2007mj,Rosa:2009ei,Pani:2012vp,Dolan:2012yt,Witek:2012tr,Brito:2014wla,Hod:2016iri,Huang:2016qnk,East:2017mrj}). In particular, quite recently, interesting progress has been achieved towards understanding the end state of the superradiant instability of the Kerr BH~\cite{East:2017ovw,Herdeiro:2017phl}, possible leading to an equilibrum state between the field and the (spun down) BH~\cite{Herdeiro:2014goa,Herdeiro:2015gia,Herdeiro:2016tmi}. 

A physically distinct, but equally interesting trapping mechanism is to consider the rotating BH in an  asymptotically anti-de-Sitter (AdS) spacetime. Superradiant instabilities are indeed present for rotating AdS BHs, at least when they are sufficiently small (see $e.g.$~\cite{Cardoso:2004hs,Li:2012rx,Cardoso:2013pza,Wang:2015fgp,Delice:2015zga}). An intriguing exception to this rule is the geometrically elegant three-dimensional BTZ BH~\cite{Banados:1992wn,Banados:1992gq}, which does not exhibit superradiance, for the simplest Bosonic field model and simplest (field vanishing, $i.e.$, Dirichlet) boundary conditions of the Bosonic field at infinity~\cite{Ortiz:2011wd}.

Dirichlet-type boundary conditions are not, however, the most general boundary conditions that can be imposed for test fields in AdS. A more general family of possible boundary conditions is obtained by employing the physical principle that they must guarantee the vanishing of the field's energy flux at the AdS boundary (see $e.g.$~\cite{Wang:2015goa,Wang:2015fgp,Dappiaggi:2016fwc}). In a recent paper~\cite{Ferreira:2017cta}, partly motivated by~\cite{Iizuka:2015vsa}, we have considered these more general  \textit{Robin} boundary conditions (RBCs) for a scalar field in the BTZ background\footnote{See also \cite{Bussola:2017wki} for a discussion of the quantum field theory of a massive scalar field in the BTZ BH with RBCs.} and we have shown that they allow the existence of stationary clouds around the BTZ BH, $i.e.$ non-trivial scalar field configurations, at the test field level, that are neither being absorbed, nor amplified by the BH. These configurations lie at the threshold of superradiant modes, suggesting the existence of the latter. In this paper we will show that indeed, under RBCs, a scalar field can exhibit superradiance in the BTZ BH spacetime. 

Our study of the superradiance of BTZ BHs under RBCs reveals one quite peculiar feature. Using natural definitions of energy and time in this spacetime, one observes that modes that extract energy from the BH are only a subset of the modes that grow exponentially in time. We interpret this behaviour as being related to an instability of the AdS background \textit{itself}: a scalar field with these RBCs has purely (positive) imaginary frequency modes in the Poincar\'e patch of AdS$_3$, signalling exponentially growing modes, and thus a \textit{bulk} instability. 

Placing a BTZ BH in AdS$_3$ leads to an interplay between the two instabilities. Low frequency modes (as compared to the angular velocity of the horizon, $\Omega_{\mathcal{H}}$) are in principle able to extract energy and angular momentum from the BH by superradiance. However, if the RBC also triggers the bulk instability the latter may be stronger and pump more energy into the BH than superradiance is able to extract. We observe that for moderately low frequencies BH energy extraction exists, so that superradiance dominates.  But for the lowest frequencies the AdS bulk instability is always dominant, and energy is transferred into the BH; still the scalar field mode is  exponentially growing (and angular momentum is being extracted). 

This interpretation is corroborated  by the existence of \textit{two different types} of RBCs (for given set of other parameters) for which the energy flux through the horizon vanishes:
\begin{itemize}
\item[$\bullet$]  For the first boundary condition, the vanishing flux through the horizon coincides with a vanishing imaginary part of the frequency and a real part that equals the angular velocity of the horizon, $\omega=k\Omega_{\mathcal{H}}$ (where $k$ is the azimuthal quantum number). These are the stationary clouds discussed previously in~\cite{Ferreira:2017cta} (see also~\cite{Dias:2011at,Hod:2012px,Hod:2013zza,Herdeiro:2014goa,Hod:2014baa,Benone:2014ssa,Wilson-Gerow:2015esa,Bernard:2016wqo,Sakalli:2016xoa,Hod:2016lgi,Herdeiro:2016tmi} for other discussions of analogous clouds). They occur at the threshold of the superradiant instability.
\item[$\bullet$] For the second boundary condition, however, the imaginary part of the frequency is positive and the real part of the frequency is in the (naively) superradiant regime, $\Re[\omega]<k \Omega_{\mathcal{H}}$. This supports the view that a dynamical equilibrium has been established across the horizon, with a non-vanishing energy entering the BH, due to the bulk instability, which is being precisely cancelled by the energy extracted from the BH by superradiance. The cancellation, however, does not occur at the level of angular momentum flux, which keeps being extracted from the BH.
\end{itemize}

\bigskip

The content of the present paper is as follows. In Section~\ref{section2} the BTZ geometry and Klein-Gordon equation on this geometry are briefly reviewed. In Section~\ref{section3} we discuss the boundary conditions to be employed in discussing superradiance, both at the horizon and at infinity. In Section~\ref{sec:superradiantmodes} we provide a definition of superradiant modes appropriate to the case at hand, and in Section~\ref{sec:numresults} we compute these superradiant modes numerically for several examples and discuss the results. Conclusions are presented in Section~\ref{section_conclusions}. The existence of unstable modes in the Poincar\'e patch of AdS, under RBCs, is discussed in Appendix~\ref{sec:AdS}. Throughout the paper we employ natural units in which $c = G_{\rm N} = 1$ and a metric with signature $(-++)$.

%%%%%%%%%%%%%%%%%%%%%%%%%%%%%%
%%%%%%%%%%%%%%%%%%%%%%%%%%%%%%
\section{Scalar field in the BTZ BH}
\label{section2}
%%%%%%%%%%%%%%%%%%%%%%%%%%%%%%
%%%%%%%%%%%%%%%%%%%%%%%%%%%%%%

%%%%%%%%%%%%%%%%%%%%%%%%%%%%%%
\subsection{BTZ BH}
%%%%%%%%%%%%%%%%%%%%%%%%%%%%%%

The metric of a non-extremal BTZ BH in Schwarzschild-like coordinates is
\begin{align}
\dd s^2 = - N(r)^2 \dd t^2 + \frac{\dd r^2}{N(r)^2} + r^2 \left(\dd \varphi + N^{\varphi}(r) \dd t \right)^2 \, ,
\end{align}
where
\begin{align}
N(r)^2 = - M + \frac{r^2}{\ell^2} + \frac{J^2}{4r^2} \, , \quad
N^{\varphi}(r) = - \frac{J}{2r^2} \, ,
\end{align}
$M$ is the mass of the BH and $J$ is its angular momentum. Without loss of generality we assume that $J>0$, whereas the non-extremal condition reads $J < M\ell$. This BH solution has an event horizon at $r=r_+$ and an inner horizon at $r=r_-$, with
\begin{equation}
r_{\pm} = \frac{\ell^2}{2} \left(M \pm \sqrt{M^2 - \frac{J^2}{\ell^2}}\right) \, ,
\end{equation}
and there is an ergoregion for $r_+ < r < r_{\rm erg} = \ell M$. However, there is no speed of light surface, that is, a surface in the exterior region for which the Killing generator of the horizon, $\chi = \partial_t + \Omega_{\mathcal{H}} \partial_{\varphi}$, becomes null outside the horizon, where
\begin{equation}
\Omega_{\mathcal{H}} = \frac{r_-}{\ell r_+} 
\end{equation}
is the angular velocity of the horizon. For a non-extremal BH, $\chi$ is always timelike outside the horizon. This is a key distinction with Kerr spacetime, for which there is always a speed of light surface, on which the event horizon generator becomes null outside the horizon. Such surface intersects the Kerr ergosphere ($i.e.$ the boundary of the ergoregion) for a sufficiently high rotation parameter~\cite{Gibbons:2008zi}.

%%%%%%%%%%%%%%%%%%%%%%%%%%%%%%
\subsection{Klein-Gordon equation}
%%%%%%%%%%%%%%%%%%%%%%%%%%%%%%

We consider a massive scalar field $\Phi$ which satisfies the Klein-Gordon equation,
\begin{equation}
	\left(\Box-\frac{\mu^2}{\ell^2}\right) \Phi = 0 \, ,
\end{equation}
where $\mu^2/\ell^2$ is the effective squared mass, which may include a coupling to the curvature as the Ricci scalar $R=-6/\ell^2$ is a constant. We assume that the Breitenlohner-Freedman bound, $\mu^2 \geq -1$, is satisfied \cite{Breitenlohner:1982jf}.

Taking the ansatz $\Phi(t,r,\varphi) = e^{-i\omega t + ik\varphi} \phi(r)$, with $\omega \in \bC$ and $k \in \mathbb{Z}$, and introducing a new radial coordinate
\begin{equation}
z = \frac{r^2-r_+^2}{r^2-r_-^2} \, ,
\end{equation}
we obtain two linearly independent solutions for $\phi(z)$ when\footnote{In this paper, we will only consider the case $-1 < \mu^2 < 0$; then, RBCs can be imposed at spatial infinity. The special case $\mu^2 = -1$ needs to studied separately and we will not pursue it in this paper.} $\mu^2 \neq n^2-1, \ n \in \mathbb{N}_0$ (labelled D = Dirichlet and N = Neumann, as explained  below),
\begin{align}
	\phi^{\rm (D)}(z) &= z^{\alpha} (1-z)^{\beta} \notag \\
	&\quad \times F(a,b;a+b+1-c;1-z) \, , \label{eq:solD} \\
	\phi^{\rm (N)}(z) &= z^{\alpha} (1-z)^{1-\beta} \notag \\
	&\quad \times F(c-a,c-b;c-a-b+1;1-z) \, , \label{eq:solN}
\end{align}
where
\begin{gather*}
	\alpha \equiv  - i \, \frac{\ell^2 r_+ }{2 (r_+^2-r_-^2)}(\omega-k\Omega_{\mathcal{H}}) \, , \ \ \
	\beta \equiv \frac{1}{2}\left(1+\sqrt{1+\mu^2}\right) \, , \\
	a \equiv \beta - i\ell \, \frac{\omega\ell+k}{2(r_+ +r_-)} \, , \qquad
	b \equiv \beta - i\ell \, \frac{\omega\ell-k}{2(r_+ - r_-)} \, , \\
	c \equiv 1 + 2\alpha \, ,
\end{gather*}
and $F$ is the Gaussian hypergeometric function. Note that the system has the symmetry $(\omega,k) \to (-\omega^*,-k)$ and, hence, we take $k \geqslant 0$ without loss of generality.

%%%%%%%%%%%%%%%%%%%%%%%%%%%%%%
%%%%%%%%%%%%%%%%%%%%%%%%%%%%%%
\section{Boundary conditions}
\label{section3}
%%%%%%%%%%%%%%%%%%%%%%%%%%%%%%
%%%%%%%%%%%%%%%%%%%%%%%%%%%%%%

In this section, we introduce the physical boundary conditions that the mode solutions of the Klein-Gordon equation in the exterior region of the BTZ BH must satisfy at both the horizon and infinity.

%%%%%%%%%%%%%%%%%%%%%%%%%%%%%%
\subsection{At the horizon}
%%%%%%%%%%%%%%%%%%%%%%%%%%%%%%

In order for superradiance to occur, we require ``ingoing'' boundary conditions at the horizon. To impose these conditions, it is convenient to consider another set of linearly independent solutions,
\begin{align} \label{eq:horizonbasis}
\phi(z) &= A \, z^{\alpha} (1-z)^{\beta} F(a,b;c;z) 
+ B \, z^{-\alpha} (1-z)^{\beta} \notag \\
&\quad \times F(a-c+1,b-c+1;2-c;z) \, .
\end{align}
The first term corresponds to ingoing modes at the horizon $z=0$, whereas the second term corresponds to outgoing modes. Hence, we set $B=0$.

%%%%%%%%%%%%%%%%%%%%%%%%%%%%%%
\subsection{At infinity}
%%%%%%%%%%%%%%%%%%%%%%%%%%%%%%

In Refs.~\cite{Ferreira:2017cta,Bussola:2017wki}, when dealing with scalar fields around the BTZ BH, we regarded the spacetime as an isolated system, which is achieved by imposing RBCs at spatial infinity when the mass parameter of the scalar field is such that $-1 < \mu^2 < 0$. For this range of the scalar field mass, both solutions \eqref{eq:solD} and \eqref{eq:solN} are square-integrable near infinity,
\begin{equation}
\int^{\infty} \dd r \sqrt{-g} \, g^{tt} \big|\phi^{\rm (D),(N)}(r)\big| < \infty \, .
\end{equation}
When $\mu^2 \geqslant 0$ only $\phi^{\rm (D)}$ is square-integrable near infinity and no RBCs need be imposed at infinity.

Therefore, for $-1 < \mu^2 < 0$, the scalar field may be written as
\begin{equation} \label{eq:RBC}
\phi(z) \propto \cos(\zeta) \phi^{\rm (D)}(z) + \sin(\zeta) \phi^{\rm (N)}(z) \, ,
\end{equation}
where $\zeta \in [0,\pi)$ parametrizes the RBCs. The standard Dirichlet boundary conditions correspond to $\zeta=0$ and we may denote by Neumann boundary conditions the case $\zeta=\frac{\pi}{2}$. However, this choice is not unique, as we can take $\phi^{\rm (N)}$ to be any other linearly independent solution.

%%%%%%%%%%%%%%%%%%%%%%%%%%%%%%
%%%%%%%%%%%%%%%%%%%%%%%%%%%%%%
\section{Superradiant mode solutions}
\label{sec:superradiantmodes}
%%%%%%%%%%%%%%%%%%%%%%%%%%%%%%
%%%%%%%%%%%%%%%%%%%%%%%%%%%%%%

Imposing boundary conditions at both the horizon and infinity selects a discrete set of mode solutions, whose frequencies we compute in this section. In constrast to the case in which Dirichlet boundary conditions are imposed at infinity, these frequencies can only be obtained numerically and we present the numerical results in the next section. After obtaining these mode solutions, we identify those which are superradiant by computing their energy flux across the horizon.

%%%%%%%%%%%%%%%%%%%%%%%%%%%%%%
\subsection{Computation of the modes frequencies}
\label{sec:eigenfreq}
%%%%%%%%%%%%%%%%%%%%%%%%%%%%%%

Let us focus on the case $-1 < \mu^2 < 0$ for which RBCs are imposed at infinity. If we perform the transformation $z \to 1-z$ of the hypergeometric function \cite{NIST} to \eqref{eq:horizonbasis}, we obtain
\begin{align}
\phi(z) &= A \, \Gamma(c) \left[ \frac{\Gamma(c-a-b)}{\Gamma(c-a)\Gamma(c-b)} \phi^{\rm (D)}(z) \right. \notag \\
&\qquad\qquad\; \left. + \frac{\Gamma(a+b-c)}{\Gamma(a)\Gamma(b)} \phi^{\rm (N)}(z) \right] \, .
\end{align}
Comparing with \eqref{eq:RBC}, one obtains
\begin{align} \label{eq:RBCQNM}
\cos(\zeta) \frac{\Gamma(a+b-c)}{\Gamma(a)\Gamma(b)} = \sin(\zeta) \frac{\Gamma(c-a-b)}{\Gamma(c-a)\Gamma(c-b)} \, .
\end{align}
In order for this condition to be satisfied for fixed $\zeta \in [0,\pi)$, the frequencies $\omega$ must take a discrete set of values, $\omega = \omega_n(k)$, $n \in \mathbb{N}_0$. Since the AdS geometry is a confining geometry and these frequencies will in general be complex-valued, one can regard these modes as \textit{quasi-bound} state modes.

\begin{figure*}[t!]
	\centering
	\includegraphics[width=0.415\linewidth]{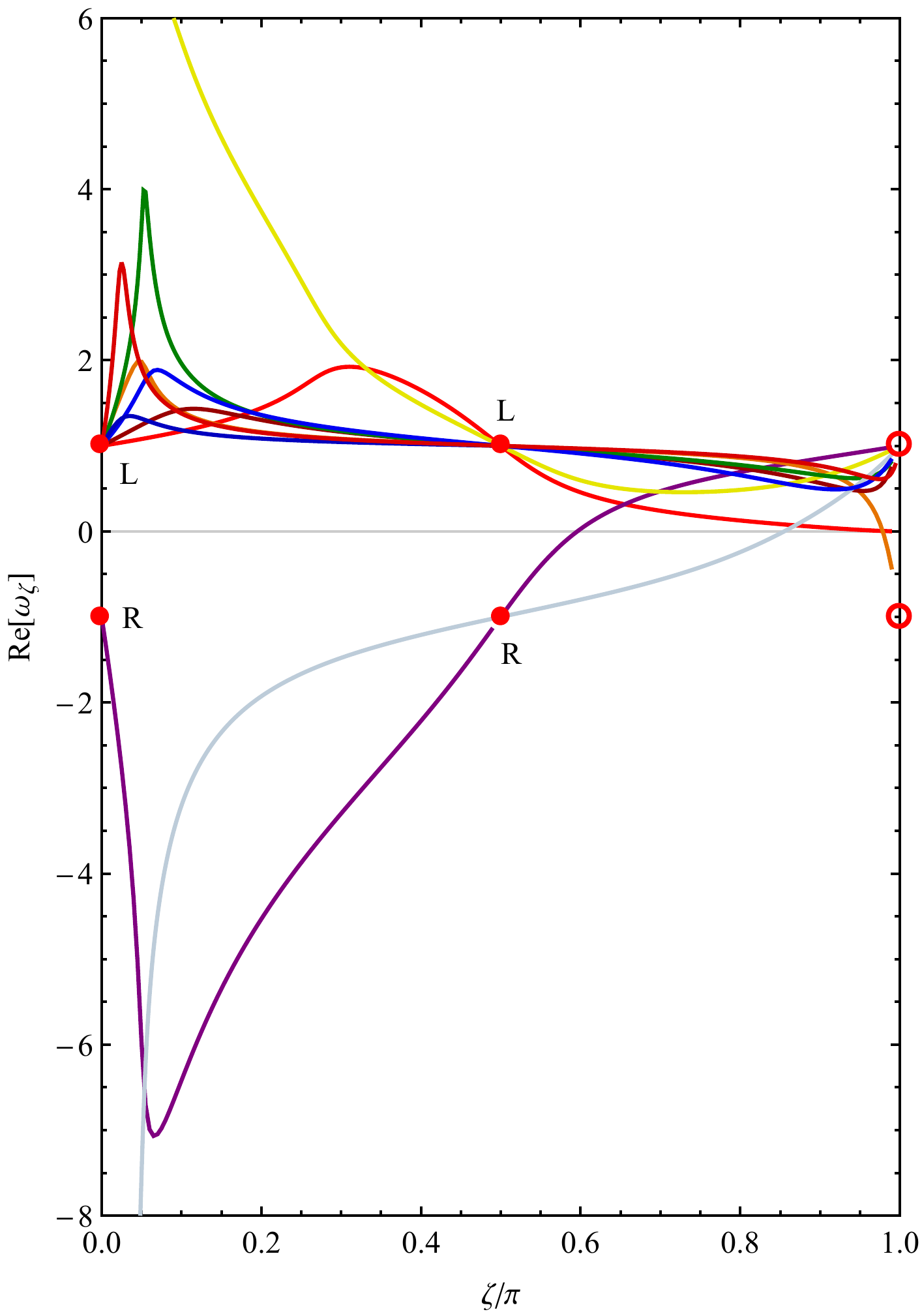} \hspace*{12ex}
	\includegraphics[width=0.415\linewidth]{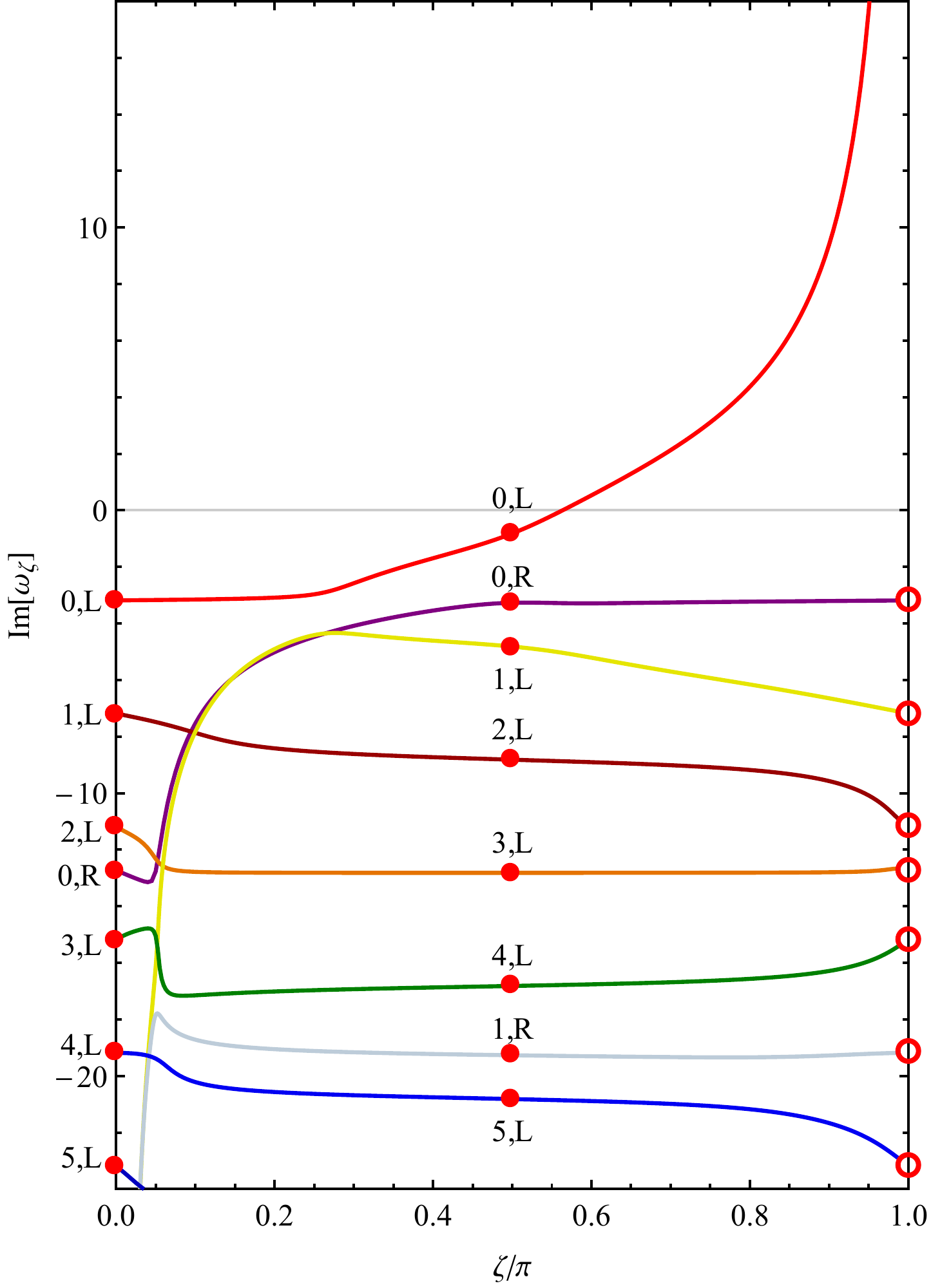}
	\caption{\label{fig:plot-BTZ-omega-1}Real and imaginary parts of some quasi-bound frequencies as a function of $\zeta/\pi$ for a BTZ BH and scalar field with $\ell=1$, $r_+=5$, $r_-=3$, $\mu^2 = -0.65$ and $k=1$ ($k \Omega_{\mathcal{H}} = 0.6$).}
\end{figure*}

This identity can be solved exactly in the cases $\zeta = 0$ (Dirichlet boundary condition) and $\zeta = \frac{\pi}{2}$ (Neumann boundary condition). In the first case, we get
\begin{align}
\omega_n^{\rm (D),L} &= \frac{k}{\ell} - i \frac{\left( r_+ - r_- \right)}{\ell^2} \left( 2n + 1 + \sqrt{1+\mu^2} \right) \, , \label{eq:omegaDL} \\
\omega_n^{\rm (D),R} &= - \frac{k}{\ell} - i \frac{\left( r_+ + r_- \right)}{\ell^2} \left( 2n + 1 + \sqrt{1+\mu^2} \right) \, , \label{eq:omegaDR}
\end{align}
with $n \in \mathbb{N}_0$, and in the second case
\begin{align}
\omega_n^{\rm (N),L} &= \frac{k}{\ell} - i \frac{\left( r_+ - r_- \right)}{\ell^2} \left( 2n + 1 - \sqrt{1+\mu^2}\right) \, , \label{eq:omegaNL} \\
\omega_n^{\rm (N),R} &= - \frac{k}{\ell} - i \frac{\left( r_+ + r_- \right)}{\ell^2} \left( 2n + 1 - \sqrt{1+\mu^2}\right) \, . \label{eq:omegaNR}
\end{align}
We distinguish the two sets of solutions as ``left'' (L) and ``right'' (R) frequencies, a nomenclature also used in \cite{Birmingham:2001hc} for the Dirichlet modes. For other RBCs, the condition has to be solved numerically. We will do so in Section~\ref{sec:numresults}, see Fig.~\ref{fig:plot-BTZ-omega-1}.

%%%%%%%%%%%%%%%%%%%%%%%%%%%%%%
\subsection{Energy flux across the horizon and definition of superradiant modes}
\label{sec:energyflux}
%%%%%%%%%%%%%%%%%%%%%%%%%%%%%%

In order to classify a given mode solution as superradiant, we check if the energy flows across the horizon \textit{towards the exterior region} of the BH. We note that the Wronskian arguments used in the case of asymptotically flat BHs (see $e.g.$ Section~4.4 of \cite{Brito:2015oca}) fail in asymptotically AdS BHs due to the imposition of RBCs at infinity, and thus cannot be used in the case at hand.

It is convenient to consider ingoing Eddington-Finkelstein (EF) coordinates $(v,r,\hat{\varphi})$, where
\begin{equation}
	\dd v = \dd t + \dd r_* \doteq \dd t + \frac{\dd r}{N^2} \, , \quad
	\dd \hat{\varphi} = \dd \varphi - \frac{N^{\varphi}}{N^2} \dd r \, ,
\end{equation}
where $r_*$ is the tortoise coordinate,
\begin{equation}
	r_* = \frac{1}{2(r_+^2-r_-^2)} \left[ r_+ \log \left(\frac{r-r_+}{r+r_+}\right) - r_- \log \left(\frac{r-r_-}{r+r_-}\right) \right] \, .
\end{equation}
Then, using \eqref{eq:horizonbasis}, a mode solution at the horizon is
\begin{equation}
\Phi(v,r_+,\hat{\varphi}) = \hat{A} \, e^{-i\omega v+ik\hat{\varphi}} \, ,
\end{equation}
with $\hat{A} = A \left(\frac{4r_+^2}{r_+^2-r_-^2}\right)^{\alpha}$.

The energy flux across the horizon for a mode solution as the one above, and allowing $\omega \in \bC$, is given by
\begin{align} 
\mathcal{F}_E(v) &= \int_0^{2\pi} \dd \hat{\varphi} \, r_+ \chi_{\mu} {T^{\mu}}_{\nu} k^{\nu}  \notag \\
&= F \left[\Im[\omega]^2 + \Re[\omega](\Re[\omega]-k\Omega_{\mathcal{H}}) \right] e^{2 \Im[\omega]v}  \, . \label{eq:energyflux}
\end{align}
where $F = 2\pi r_+ \hat{A}^2$, $\chi = \partial_v + \Omega_{\mathcal{H}} \partial_{\hat{\varphi}}$ is the generator of the horizon, $k = \partial_v$ and $T_{\mu\nu}$ is the stress-energy tensor of the scalar field,
\begin{equation}
T_{\mu\nu} = \partial_{(\mu} \overline{\Phi} \partial_{\nu)} \Phi - \frac{1}{2} g_{\mu\nu} \left(g^{\rho\lambda} \partial_{(\rho} \overline{\Phi} \partial_{\lambda)} \Phi + \frac{1}{2} \frac{\mu^2}{\ell^2} |\Phi|^2\right) \, .
\end{equation}
The expression \eqref{eq:energyflux} is in agreement with earlier calculations, e.g.~in \cite{Zouros:1979iw,Dolan:2007mj,Wilson-Gerow:2015esa}.

The energy flux across the horizon \eqref{eq:energyflux} is negative if $\Im[\omega]^2 + \Re[\omega](\Re[\omega]-k\Omega_{\mathcal{H}}) < 0$, in which case energy is coming out of the BH towards the exterior region. Observe this can only occur for $k\neq 0$ and $\Omega_{\mathcal{H}}\neq 0$. We will take as an operational definition of a \emph{superradiant mode} to be a mode for which the energy flux \eqref{eq:energyflux} is negative. A necessary condition for this flux to be negative is that $0 < \Re[\omega] < k \Omega_{\mathcal{H}}$, but it is by no means sufficient. If in addition $|{\Im[\omega]}| \ll {\Re[\omega]}$ we recover the usual result for the superradiance regime, $0 < \omega < k \Omega_{\mathcal{H}}$ (see $e.g.$~\cite{Brito:2015oca}).

Furthermore, note that in the case of stationary scalar clouds around the BTZ BH considered in Ref.~\cite{Ferreira:2017cta}, for which $\omega = k \Omega_{\mathcal{H}}$, the energy flux is identically zero, as required. As we will see below, these clouds exist at the threshold of the superradiance regime. 

The flux of angular momentum across the horizon can be computed in a similar fashion:
\begin{align} 
\mathcal{F}_L(v) &= - \int_0^{2\pi} \dd \hat{\varphi} \, r_+ \chi_{\mu} {T^{\mu}}_{\nu} m^{\nu}  \notag \\
&= F k \left(\Re[\omega]-k\Omega_{\mathcal{H}} \right) e^{2 \Im[\omega]v}  \, , \label{eq:angmomflux}
\end{align}
where $m = \partial_{\hat{\varphi}}$. So the angular momentum flux is always towards the exterior region if $0 < \Re[\omega] < k \Omega_{\mathcal{H}}$.

%%%%%%%%%%%%%%%%%%%%%%%%%%%%%%
%%%%%%%%%%%%%%%%%%%%%%%%%%%%%%
\section{Numerical results}
\label{sec:numresults}
%%%%%%%%%%%%%%%%%%%%%%%%%%%%%%
%%%%%%%%%%%%%%%%%%%%%%%%%%%%%%

In this section, we present the numerical results for the frequencies, energy flux and angular momentum flux of the quasi-bound state modes as functions of the RBCs.

\begin{figure*}[t!]
	\centering
	\includegraphics[width=0.35\linewidth]{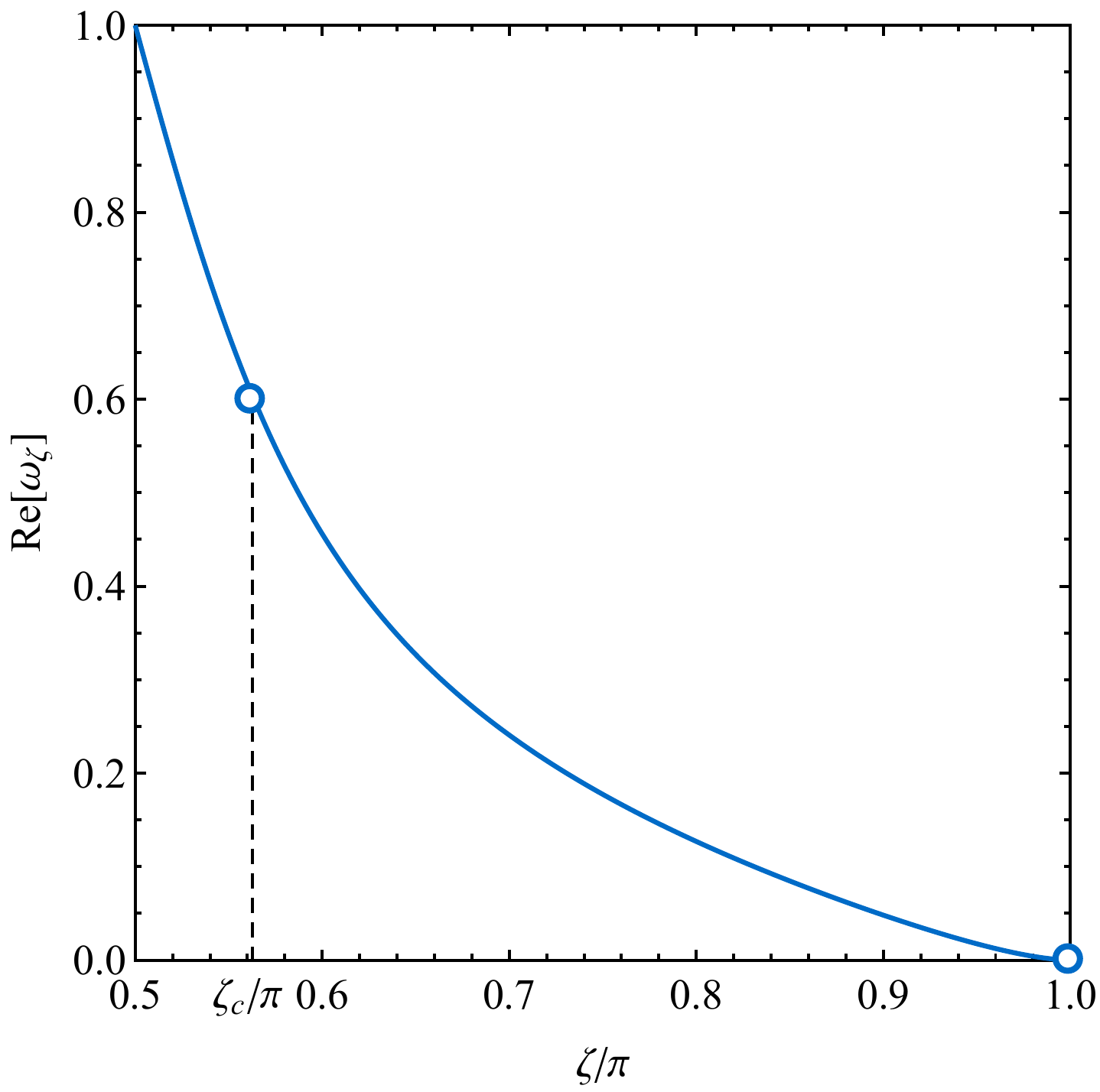} \qquad
	\includegraphics[width=0.35\linewidth]{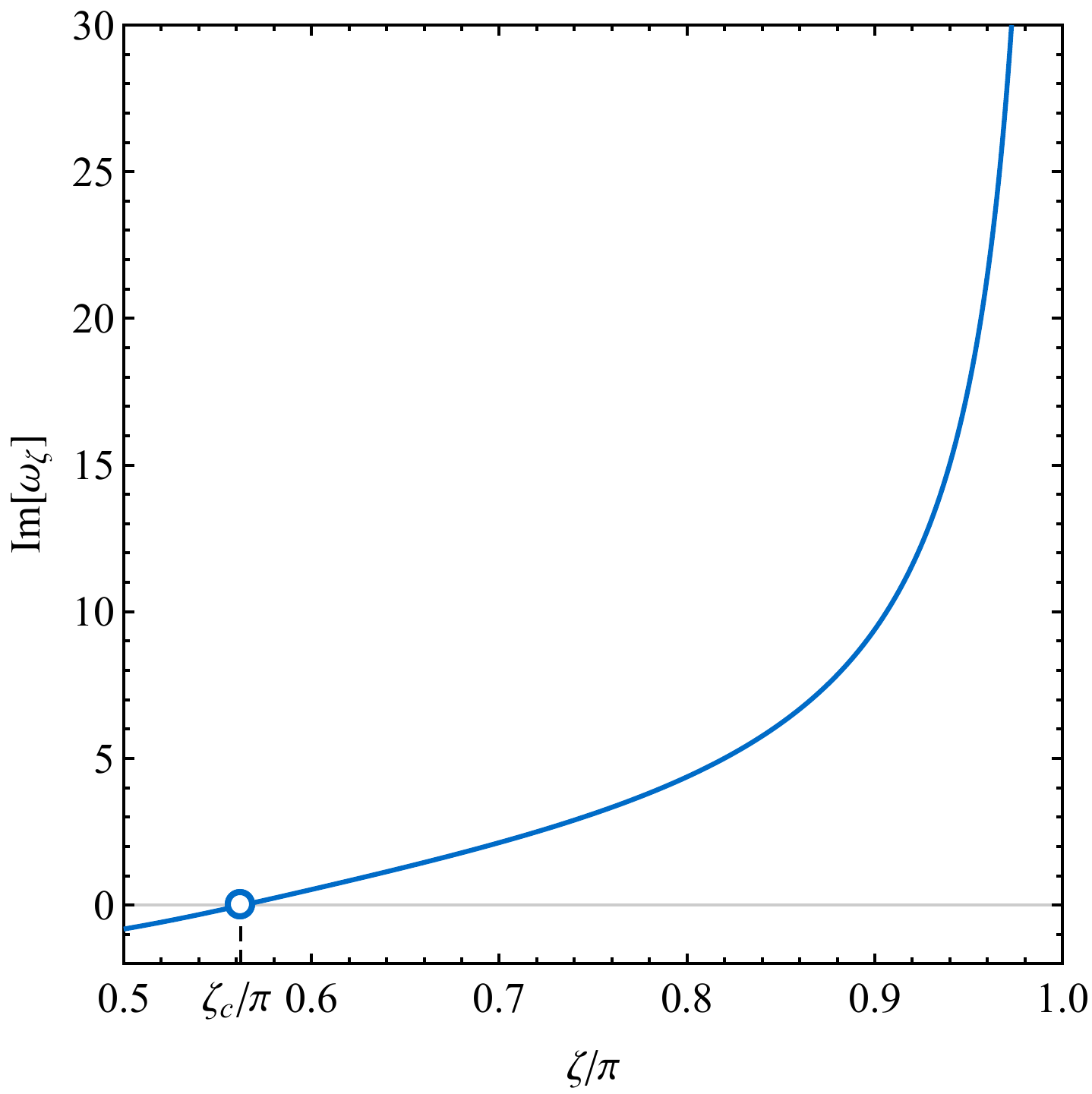} \\
	\includegraphics[width=0.35\linewidth]{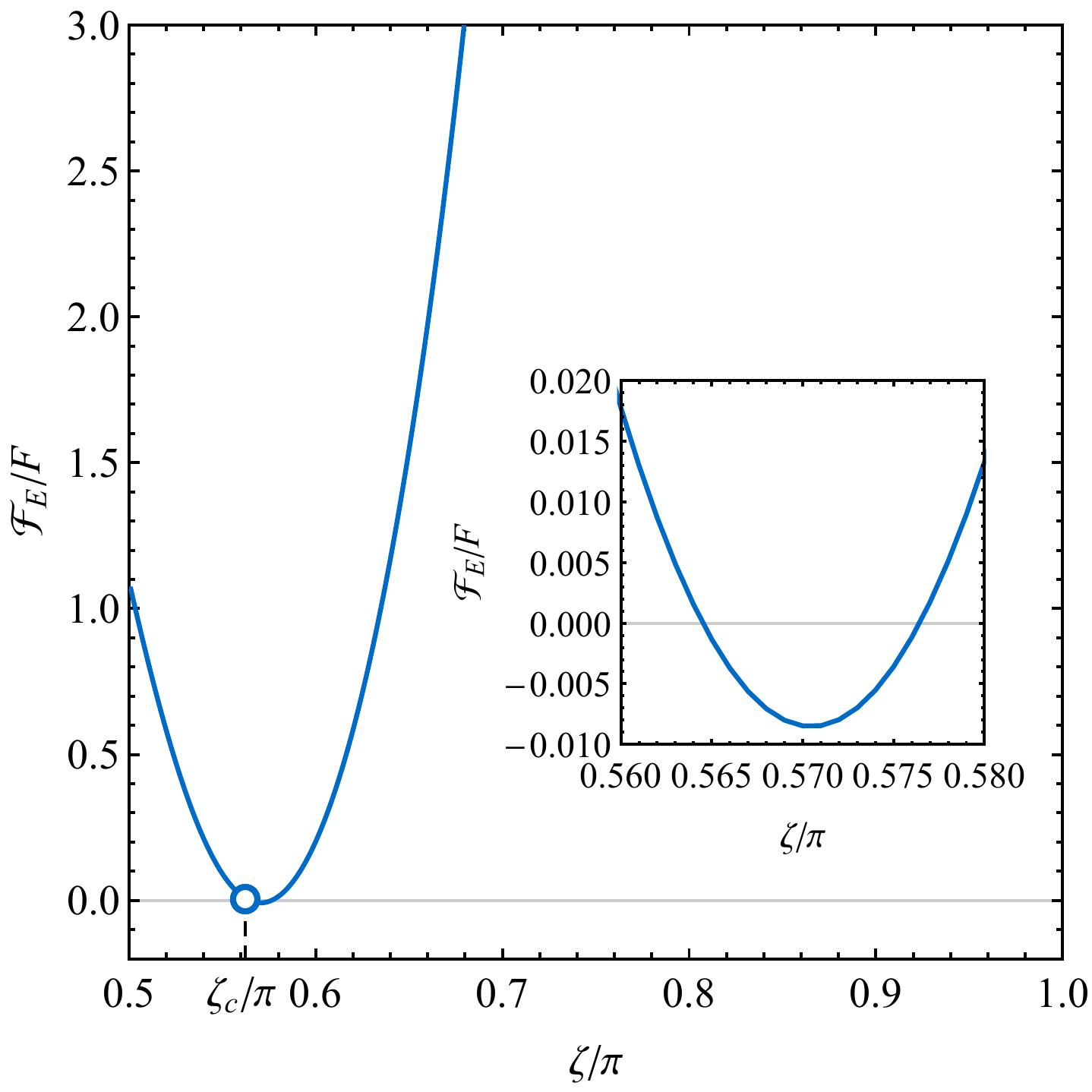} \qquad
	\includegraphics[width=0.35\linewidth]{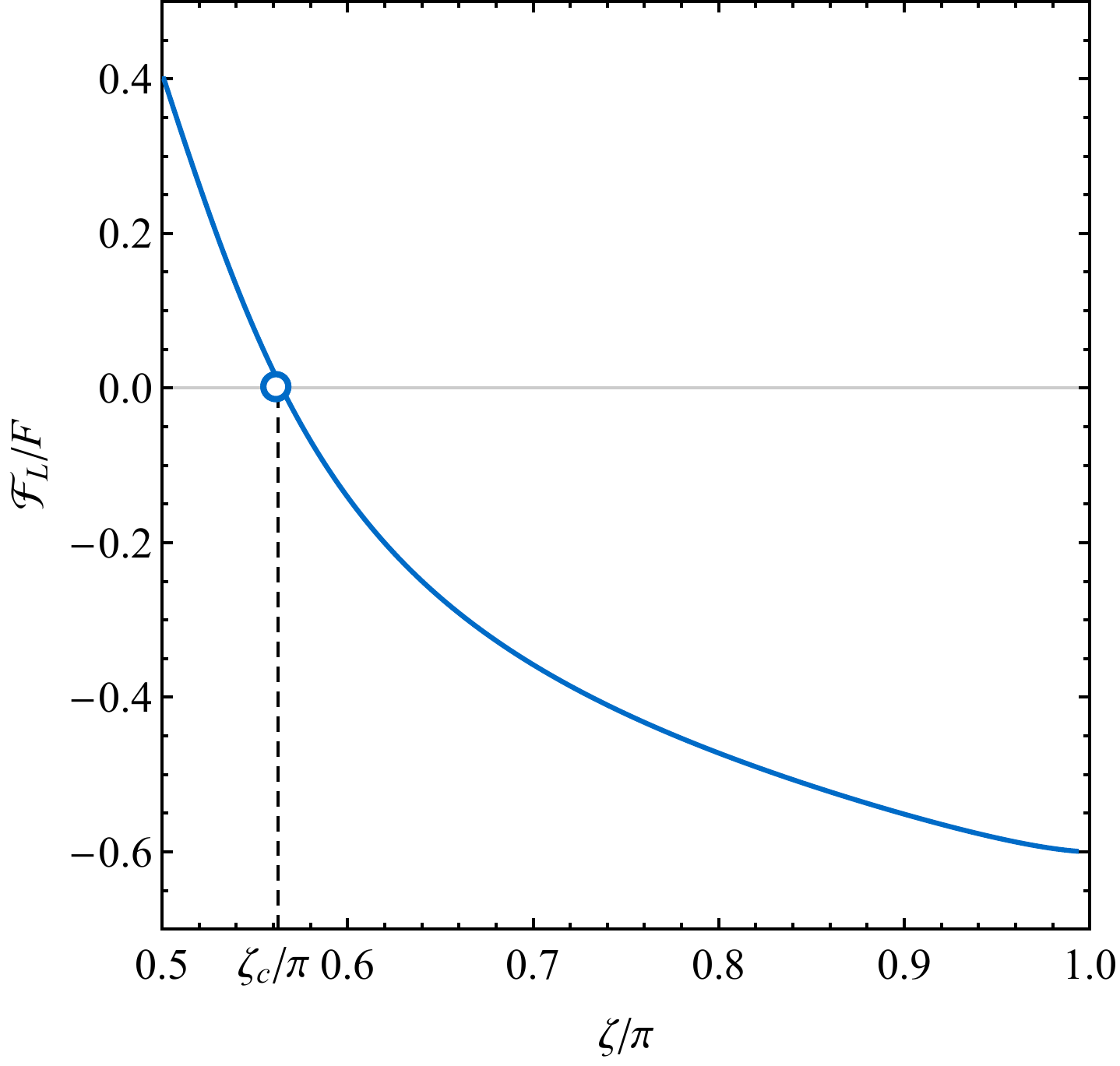}
	\caption{\label{fig:plots1}Real and imaginary parts of the frequency and fluxes of energy $\mathcal{F}_E/F$ and angular momentum $\mathcal{F}_L/F$ across the horizon for the mode 0,L as a function of $\zeta/\pi$ for a BTZ BH and scalar field with $\ell=1$, $r_+=5$, $r_-=3$, $\mu^2 = -0.65$ and $k=1$.}
\end{figure*}

\begin{figure*}[t!]
	\includegraphics[width=0.35\linewidth]{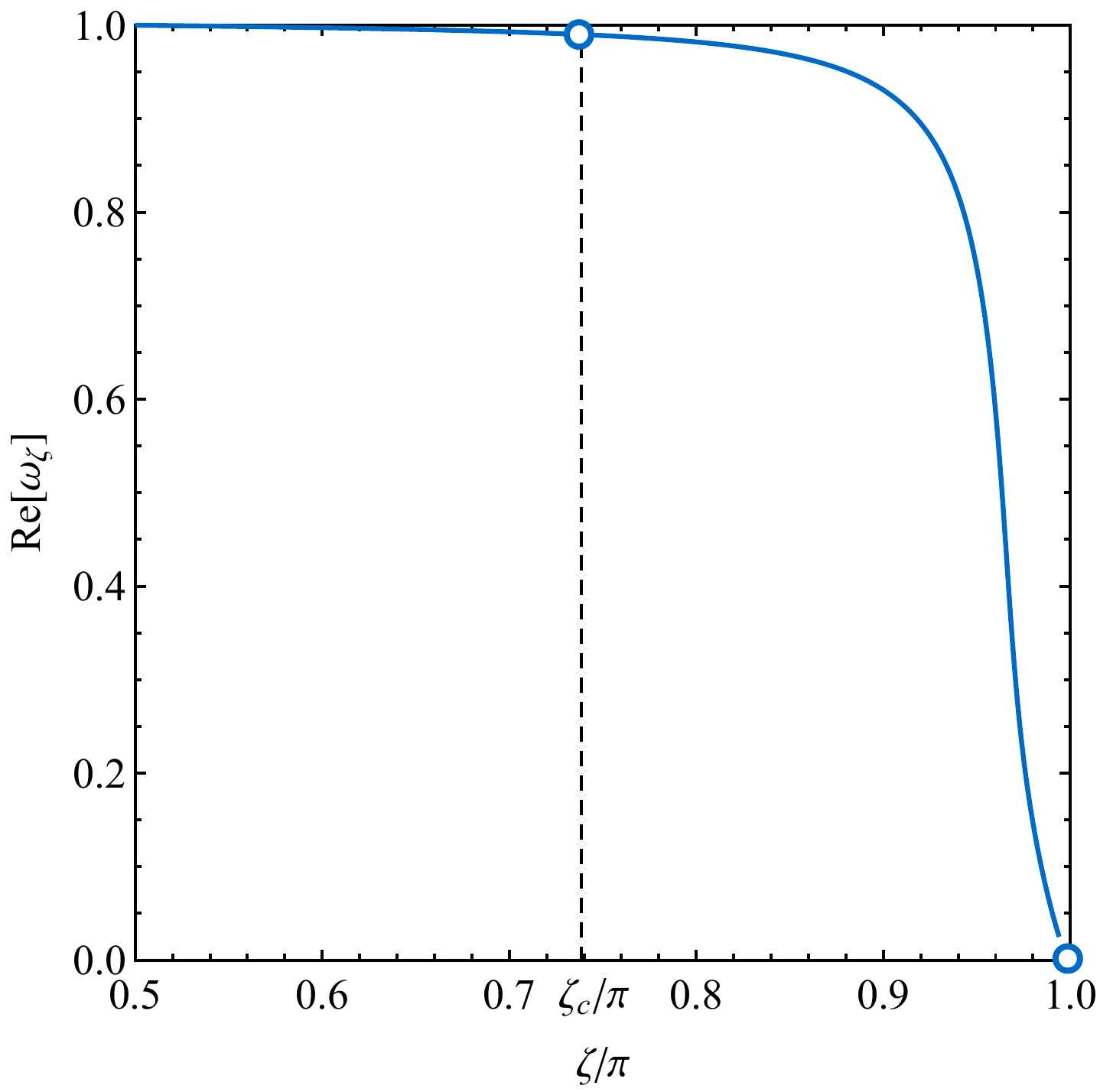} \qquad
	\includegraphics[width=0.35\linewidth]{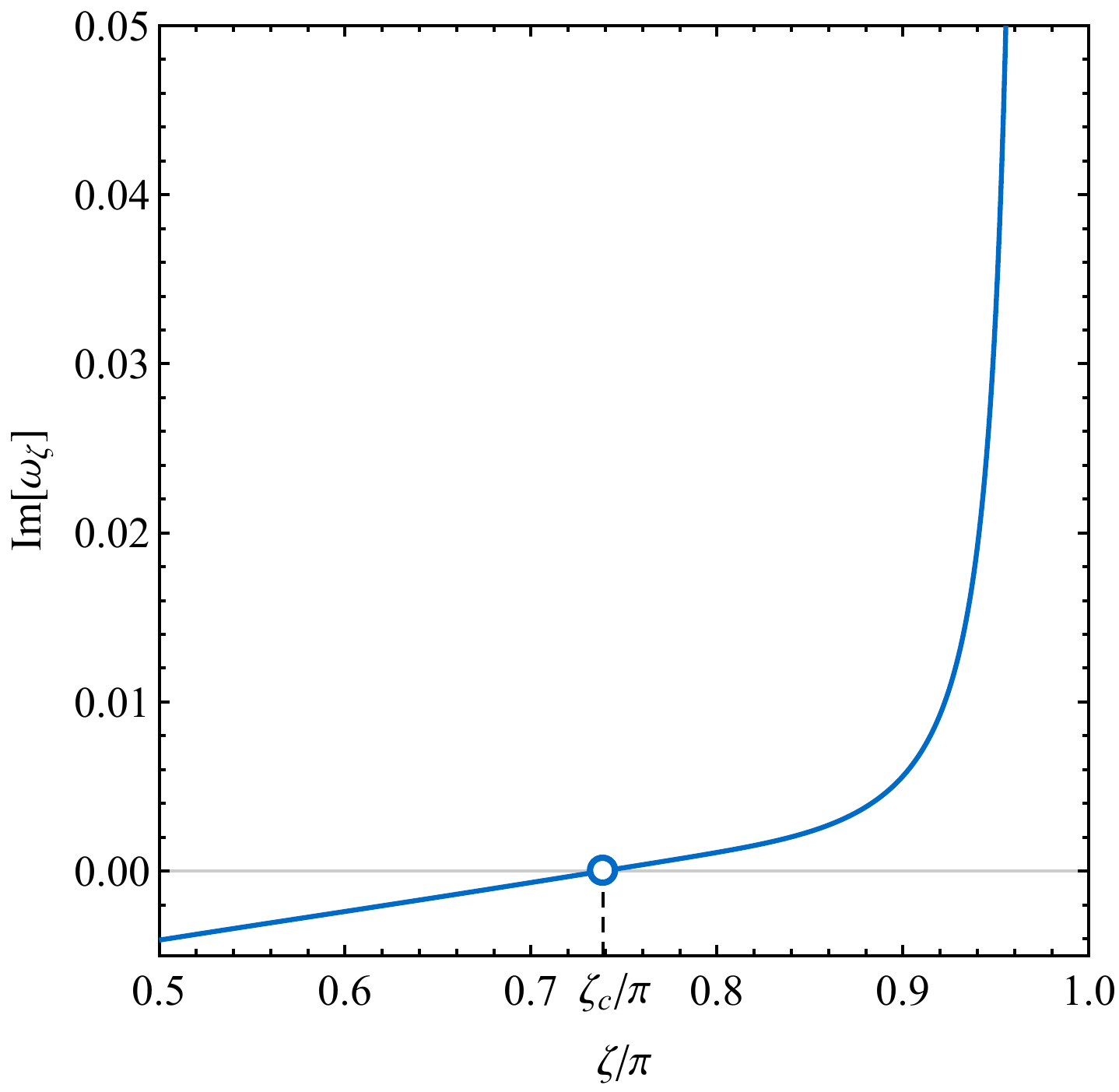} \\
	\includegraphics[width=0.35\linewidth]{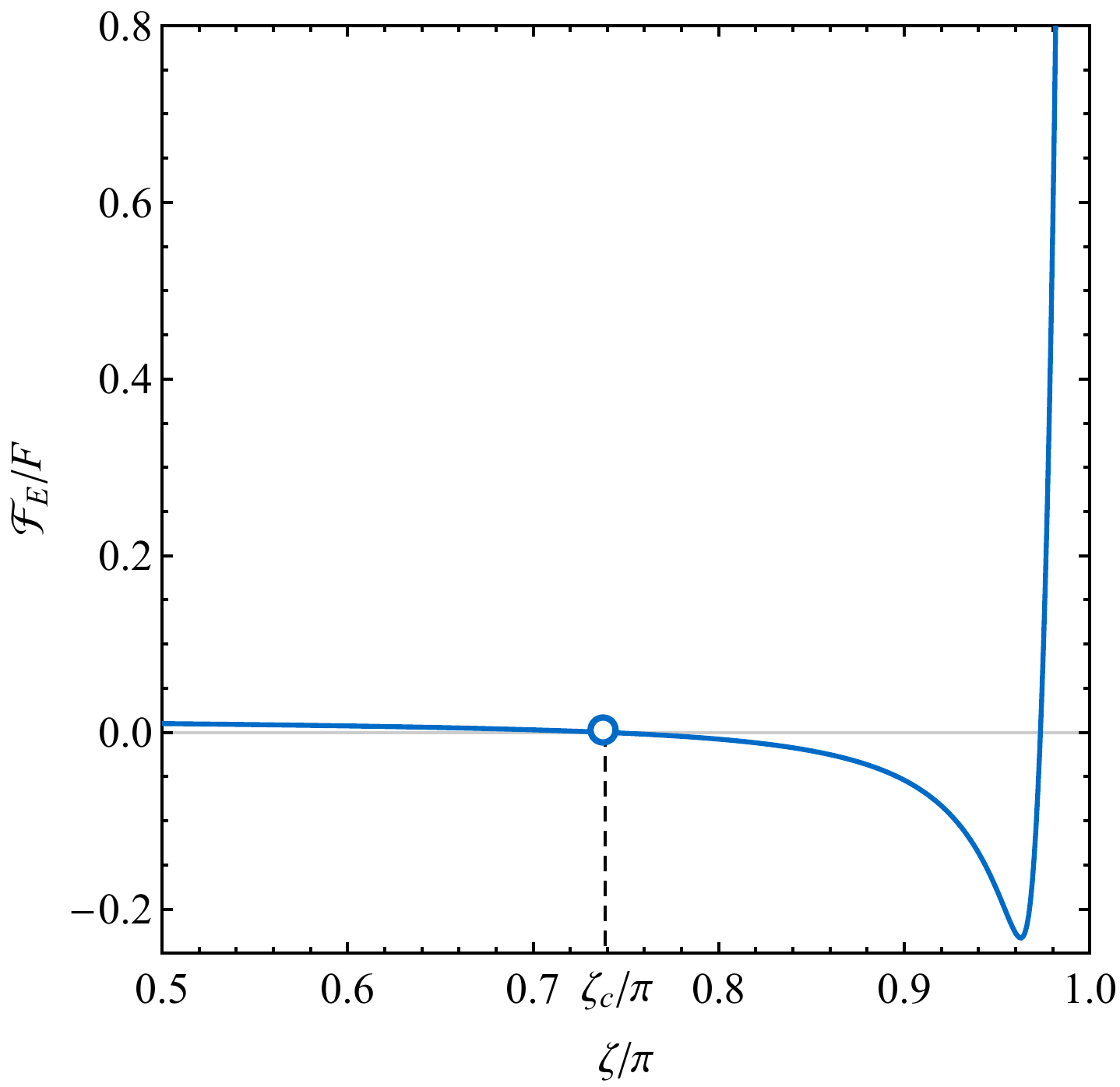} \qquad
	\includegraphics[width=0.35\linewidth]{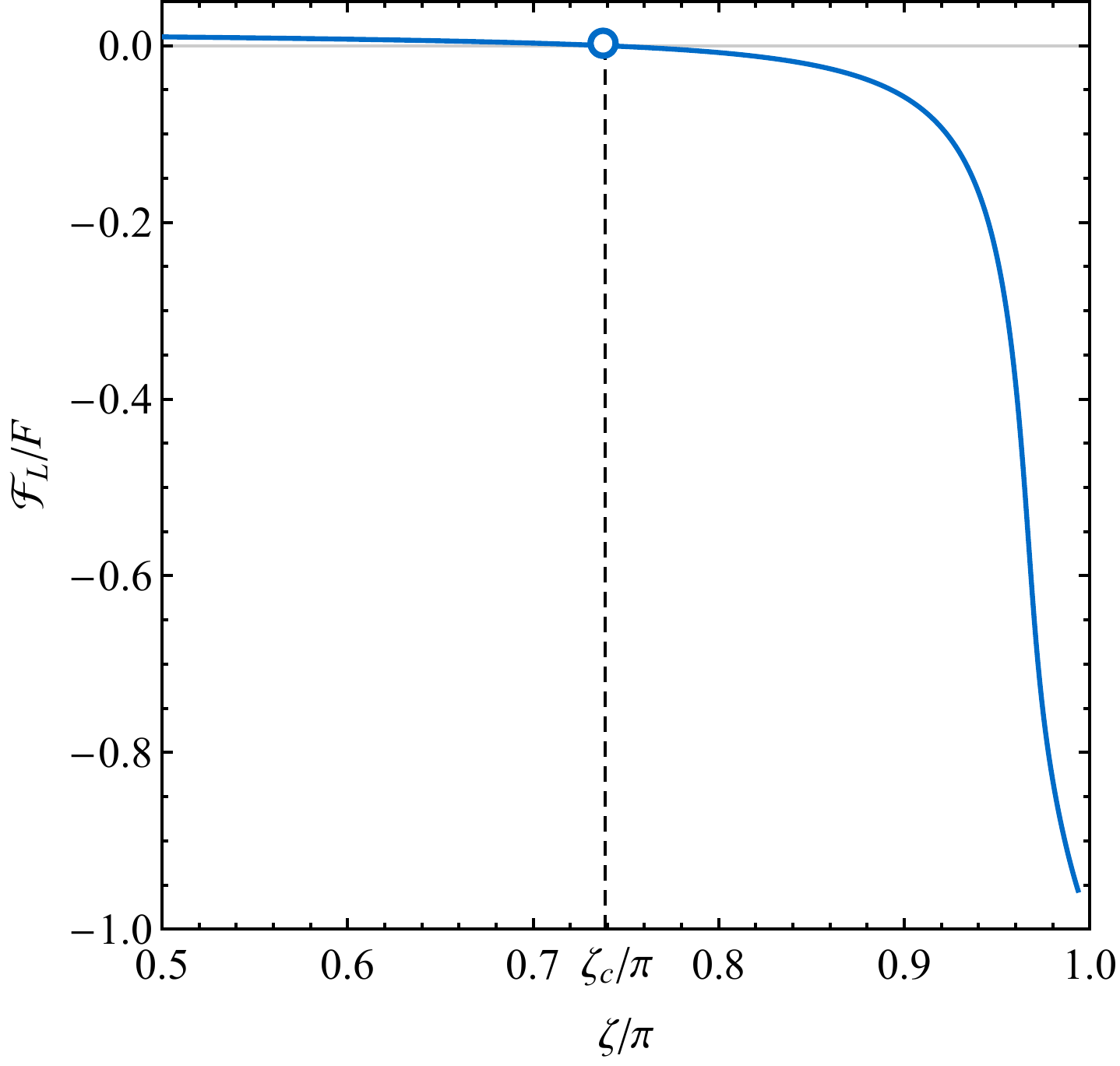}
	\caption{\label{fig:plots2}Real and imaginary parts of the frequency and fluxes of energy $\mathcal{F}_E/F$ and angular momentum $\mathcal{F}_L/F$ across the horizon for the mode 0,L as a function of $\zeta/\pi$ for a BTZ BH and scalar field with $\ell=1$, $r_+=1$, $r_-=0.99$, $\mu^2 = -0.65$ and $k=1$.}
\end{figure*}

\begin{figure*}[t!]
	\includegraphics[width=0.36\linewidth]{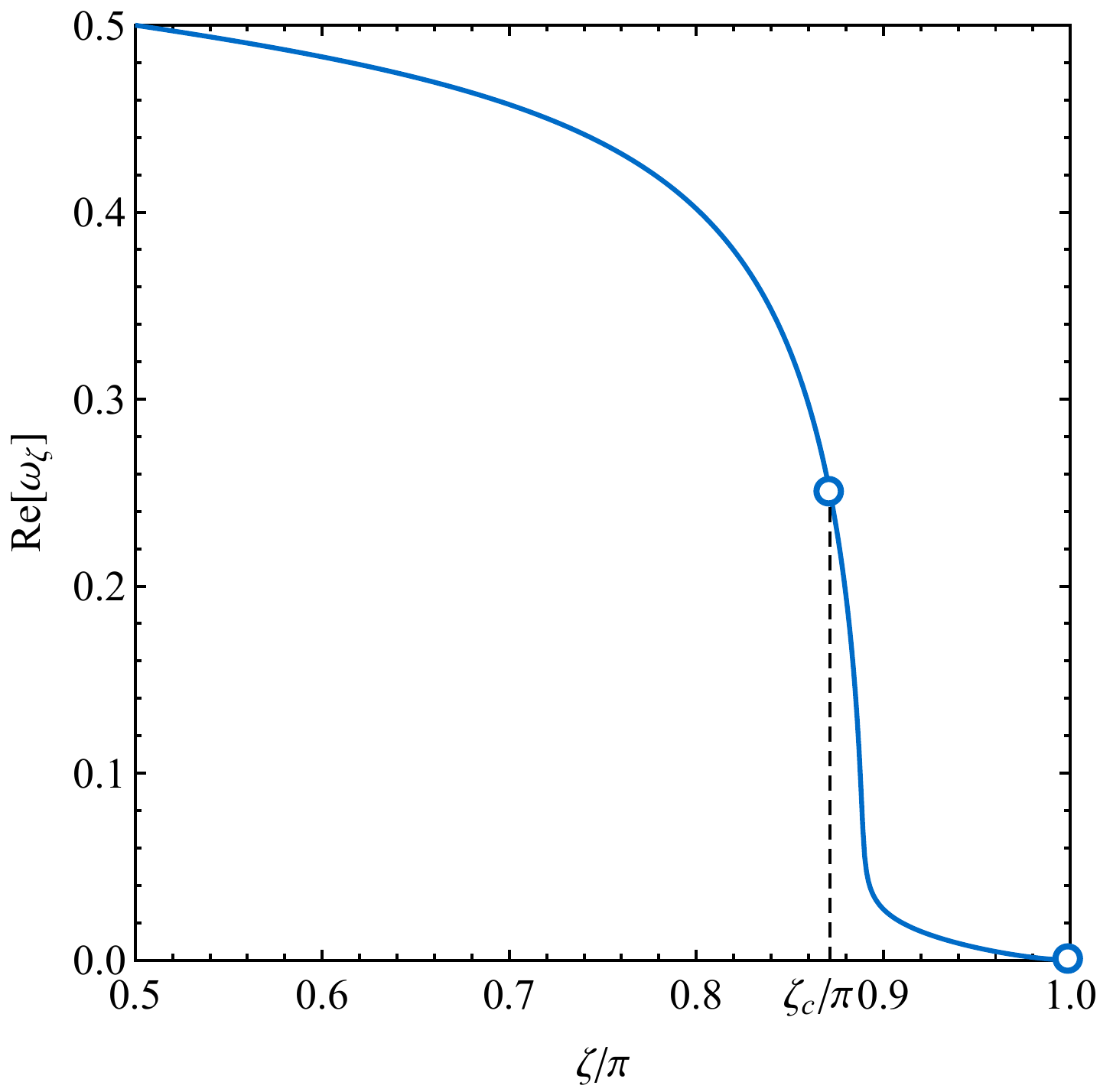} \qquad
	\includegraphics[width=0.36\linewidth]{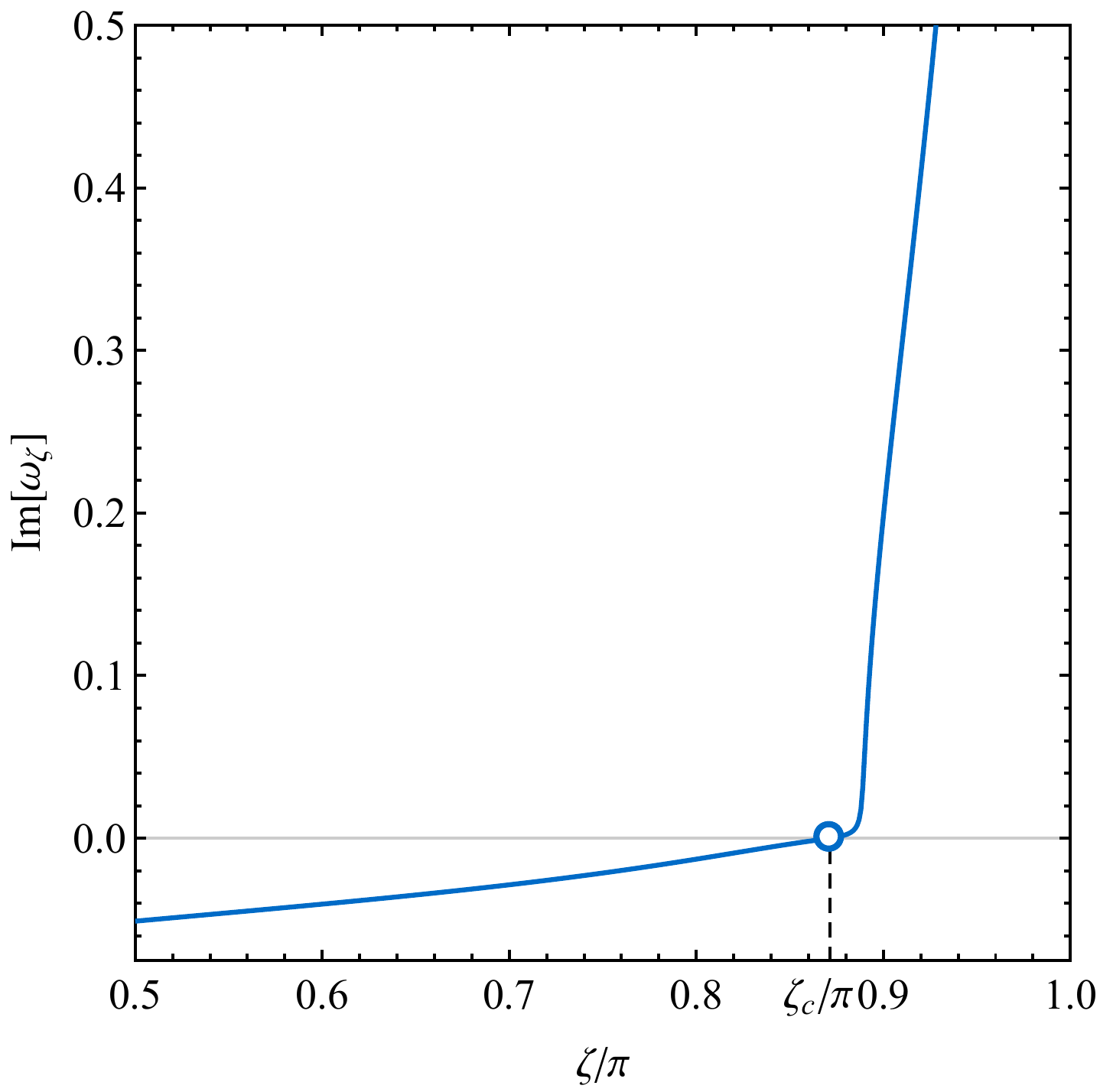} \\
	\includegraphics[width=0.36\linewidth]{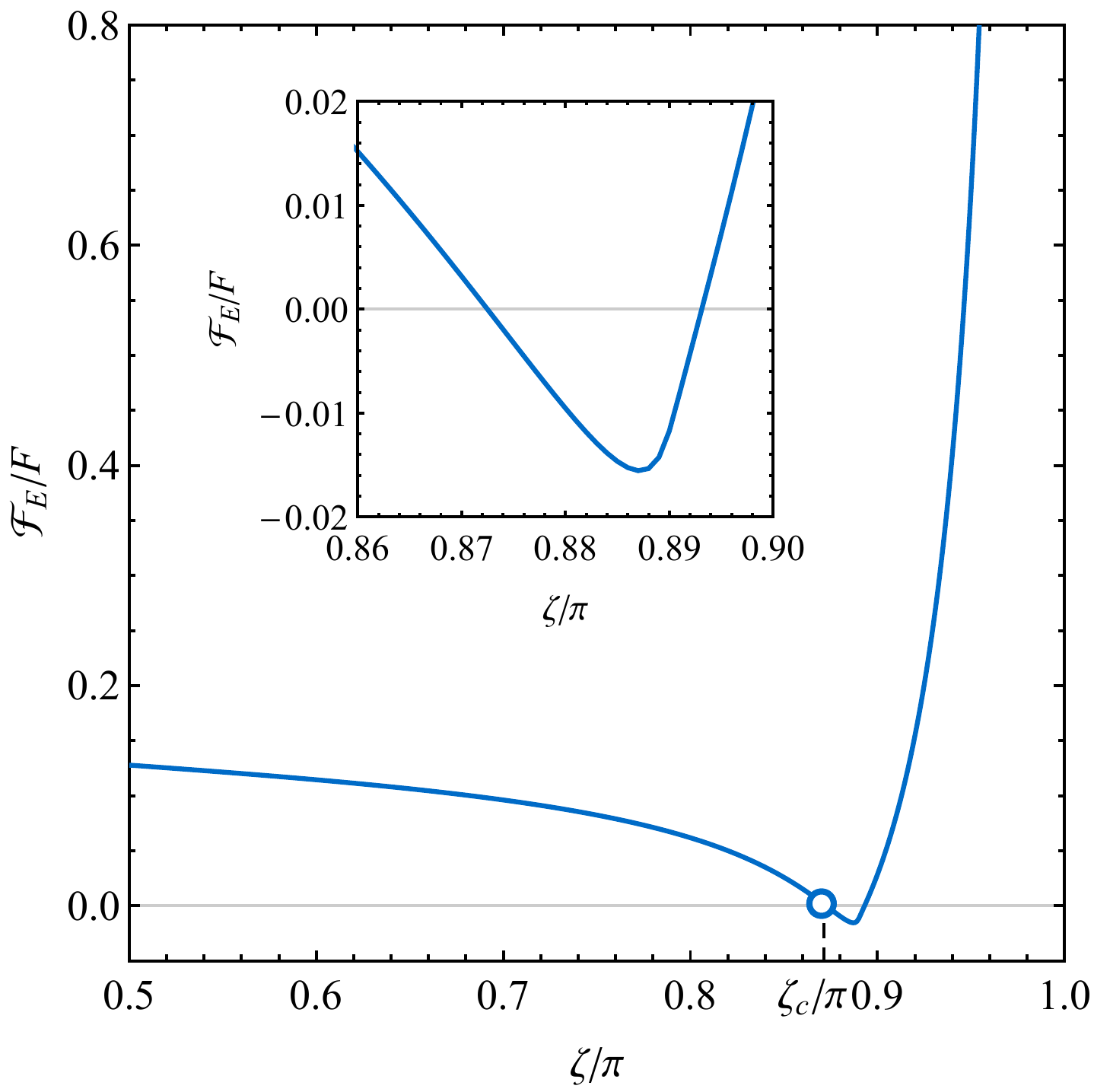} \qquad
	\includegraphics[width=0.36\linewidth]{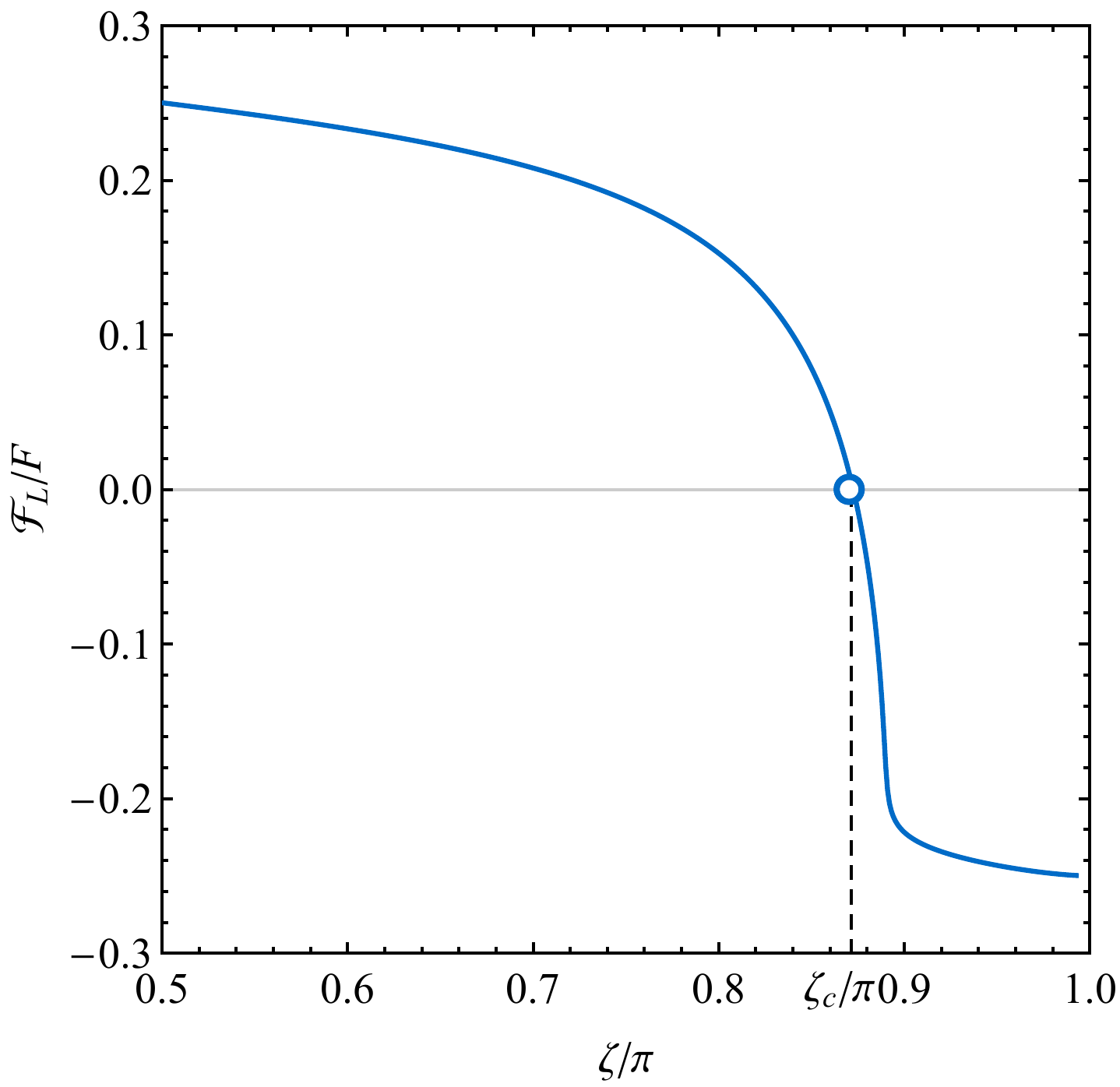}
	\caption{\label{fig:plots3}Real and imaginary parts of the frequency and fluxes of energy $\mathcal{F}_E/F$ and angular momentum $\mathcal{F}_L/F$ across the horizon for the mode 0,L as a function of $\zeta/\pi$ for a BTZ BH and scalar field with $\ell=2$, $r_+=1$, $r_-=0.5$, $\mu^2 = -0.65$ and $k=1$.}
\end{figure*}

%%%%%%%%%%%%%%%%%%%%%%%%%%%%%%
\subsection{Computation of the modes frequencies}
%%%%%%%%%%%%%%%%%%%%%%%%%%%%%%

As discussed in Section~\ref{sec:eigenfreq}, the frequencies of the mode solutions satisfying the required boundary conditions at the horizon and at infinity must take a set of discrete values. Our goal is to find if any of these mode solutions is superradiant for a fixed RBC, i.e.~if for any of these mode solutions the energy flux across  the horizon (as well as the angular momentum flux) is negative.

First, we compute these quasi-bound frequencies $\omega_{\zeta}$ for each $\zeta \in [0,\pi)$ parametrizing the RBC by numerically solving Eq.~\eqref{eq:RBCQNM} for $\omega$ for a given BH configuration (fixed $\ell$, $r_+$ and $r_-$) and scalar field configuration (fixed $\mu^2$ and $k$). We plot the real and imaginary parts of these frequencies as a function of $\zeta$ for a representative example in Fig.~\ref{fig:plot-BTZ-omega-1}. The red dots correspond to quasi-bound frequencies in the case of Dirichlet ($\zeta=0$) and Neumann ($\zeta=\frac{\pi}{2}$) boundary conditions, which can also be computed analytically [see Eqs.~\eqref{eq:omegaDL}-\eqref{eq:omegaNR}] and which served as a check of the numerical computation.

%%%%%%%%%%%%%%%%%%%%%%%%%%%%%%
\subsection{Computation of the energy flux across the horizon}
%%%%%%%%%%%%%%%%%%%%%%%%%%%%%%

Having obtained the quasi-bound frequencies for all values of $\zeta \in [0,\pi)$, we want to verify if any of them is superradiant by computing their energy flux across the horizon. Thankfully, we can restrict our attention to a small set of modes using the following arguments.

A necessary condition for a mode to be superradiant is that $0 < \Re[\omega] < k \Omega_{\mathcal{H}}$. Hence, according to Eqs.~\eqref{eq:omegaDL}-\eqref{eq:omegaNR}, for Dirichlet ($\zeta=0$) and Neumann ($\zeta=\frac{\pi}{2}$) boundary conditions, none of the modes are superradiant [recall that $\Omega_{\mathcal{H}}=r_-/(\ell r_+)$]. This is in agreement with Ref.~\cite{Ortiz:2011wd} which analyzed the Dirichlet case. Also note that for both boundary conditions the frequencies of the modes are such that $\Im[\omega]<0$.

Moreover, we know from \cite{Ferreira:2017cta} that for $\zeta = \zeta_{\rm c}$, where
\begin{align} \label{eq:zetacritical}
\tan(\zeta_{\rm c}) = \frac{\Gamma\left(2\beta-1\right)\left|\Gamma\left(1-\beta+i\frac{k}{\ell r_+}\right)\right|^2}{\Gamma\left(1-2\beta\right)\left|\Gamma\left(\beta+i\frac{k}{\ell r_+}\right)\right|^2} \, ,
\end{align}
there is one mode solution with $\Re[\omega] = k \Omega_{\mathcal{H}}$ and $\Im[\omega]=0$, the stationary scalar cloud, for which the energy flux across the horizon is zero. It is easy to see that $\zeta_{\rm c} \in (\frac{\pi}{2},\pi)$, as $\beta = \frac{1}{2}\big(1+\sqrt{1+\mu^2}\big)$ and $-1 < \mu^2 < 0$. From numerical results such as those presented in Fig.~\ref{fig:plot-BTZ-omega-1}, we verify that this mode is continuously connected, as a function of $\zeta$, to the Dirichlet and Neumann $n=0$ quasi-bound state modes of type L, Eqs.~\eqref{eq:omegaDL} and \eqref{eq:omegaNL}. For fixed RBC, all the other quasi-bound frequencies are such that $\Im[\omega]<0$.

These observations, together with \eqref{eq:energyflux}, allow us to conclude that, with fixed RBC, only the mode with frequency with largest imaginary part (continuously connected with the $n=0$ modes of type L) may have \emph{negative} energy flux across the horizon and only for $\zeta$ in a range contained in $(\zeta_{\rm c},\pi)$. Hence, in the numerical computations we performed, we focused on this mode, for which we computed its frequency and energy and angular momentum fluxes across the horizon as a function of $\zeta$. In Figs.~\ref{fig:plots1}, \ref{fig:plots2} and \ref{fig:plots3}, we present such numerical results for three sets of input data.

%%%%%%%%%%%%%%%%%%%%%%%%%%%%%%
\subsection{Discussion of the results}
\label{sec_discussion}
%%%%%%%%%%%%%%%%%%%%%%%%%%%%%%

In Figs.~\ref{fig:plots1}, \ref{fig:plots2} and \ref{fig:plots3} we presented the numerical results for the frequencies and energy and angular momentum fluxes as a function of $\zeta$ parametrizing the RBC at infinity for several representative examples of BTZ BHs and scalar field configurations. These results allow us to conclude that, for a given BTZ BH and massive scalar field, there exists a range of RBCs for which there are mode solutions whose energy flux across the horizon is \emph{negative}, that is, the energy flux is towards the exterior region of the BH. According to the definition we proposed in Section~\ref{sec:energyflux}, we identify these mode solutions as superradiant modes.

More concretely, for a BTZ BH with given $\ell$, $r_+$ and $r_-$ and a scalar field mode solution with fixed $\mu^2$ and $k$, there is a range of RBCs with $\zeta \in (\zeta_{\rm c},\zeta') \subset (\frac{\pi}{2},\pi)$ for which the energy (and angular momentum) flux across the horizon \eqref{eq:energyflux} is negative.

It is important to note that, contrarily to the intuition coming from asymptotically flat, rotating BHs, it is \emph{not} the case that all modes with $\Im[\omega] > 0$ are superradiant. For $\zeta \in (\zeta',\pi)$ the aforementioned mode solution has $\Im[\omega] > 0$ but the energy flux across the horizon is positive and, therefore, it is not superradiant.

However, we see from the numerical results that for these mode solutions $\Im[\omega]$ appears to diverge as $\zeta \to \pi^-$. To see this, rewrite \eqref{eq:RBCQNM} as
\begin{align} \label{eq:tandetailed}
\tan(\zeta) &= \frac{\Gamma\left(\sqrt{1+\mu^2}\right) \Gamma\left(\frac{1}{2}-\frac{1}{2}\sqrt{1+\mu^2}-\frac{i\ell}{2}\frac{\omega\ell-k}{r_+-r-}\right)}{\Gamma\left({-\sqrt{1+\mu^2}}\right)\Gamma\left(\frac{1}{2}+\frac{1}{2}\sqrt{1+\mu^2}-\frac{i\ell}{2}\frac{\omega\ell-k}{r_+-r-}\right)} \notag \\
&\quad \times \frac{\Gamma\left(\frac{1}{2}-\frac{1}{2}\sqrt{1+\mu^2}-\frac{i\ell}{2}\frac{\omega\ell+k}{r_++r-}\right)}{\Gamma\left(\frac{1}{2}+\frac{1}{2}\sqrt{1+\mu^2}-\frac{i\ell}{2}\frac{\omega\ell+k}{r_++r-}\right)} \, .
\end{align}
Note that the reciprocal gamma function $1/\Gamma(z)$ is an entire function of $z$ which has zeros at $z \in \mathbb{Z}_0^-$ [which correspond to the Dirichlet frequencies \eqref{eq:omegaDL} and \eqref{eq:omegaDR}]. Moreover, it tends to zero when $\Re[z] \to + \infty$. In terms of the expression above, this implies that $\Im[\omega] \to +\infty$ as $\zeta \to \pi^-$. This is indeed what is observed numerically. 

This divergence of the imaginary part of the frequency also occurs for ``bound" state mode solutions of the massive Klein-Gordon equation in the Poincar\'e patch of AdS${}_3$ for a subclass of RBCs with $\zeta$ between some value and $\pi$, as described in Appendix~\ref{sec:AdS}. This indicates that the behavior of the frequency in the BH spacetime for RBCs with $\zeta$ just below $\pi$ is due to the AdS asymptotics of the spacetime, which overshadows the superradiance effect caused by the rotating nature of the BH. This also explains why not all modes with $\Im[\omega] > 0$ are superradiant; only those with RBCs such that $\zeta \in (\zeta_{\rm c},\zeta')$ for which the energy flux across the horizon is negative are superradiant. When $\zeta \in (\zeta', \pi)$, the energy flux is positive and the growth of the mode is caused by the bulk AdS instability.

%%%%%%%%%%%%%%%%%%%%%%%%%%%%%%
%%%%%%%%%%%%%%%%%%%%%%%%%%%%%%
\section{Conclusions}
\label{section_conclusions}
%%%%%%%%%%%%%%%%%%%%%%%%%%%%%%
%%%%%%%%%%%%%%%%%%%%%%%%%%%%%%

In this paper we have shown that, for certain choices of RBCs at the conformal boundary, there exist superradiant modes of massive scalar fields in the background of a rotating BTZ BH. These choices do not include the commonly used Dirichlet boundary conditions, which explains why the phenomenon of superradiance in the BTZ BH has not been thoroughly explored in the literature so far. Moreover, as anticipated in \cite{Ferreira:2017cta}, the superradiance regime for the massive scalar field has its onset with the stationary scalar clouds which, for fixed background and field parameters, can only exist for RBC parametrized by $\zeta = \zeta_{\rm c}$ given by \eqref{eq:zetacritical}.

The AdS asymptotics, together with the RBCs and the particularities of the BTZ BH, make the overall picture on the superradiant modes significantly different from the one for asymptotically flat BHs such as the Kerr BH. In particular, we have shown that \emph{not} all mode solutions growing exponentially in time (equivalently, with frequencies with positive imaginary part) are superradiant, that is, have energy flux across the horizon towards the exterior region. In fact, for fixed spacetime and field parameters, we see two competing regimes:
\begin{itemize}
\item For RBCs with $\zeta \in (\zeta_{\rm c},\zeta') \subset (\frac{\pi}{2},\pi)$ there exists a mode solution whose frequency has positive imaginary part and whose energy (and angular momentum) flux across the horizon is \emph{negative} --- a superradiant mode.
\item For RBCs with $\zeta \in (\zeta',\pi)$ there exists a mode solution whose frequency has positive imaginary part and whose energy flux across the horizon is \emph{positive} --- an unstable mode due to the AdS asymptotics. Still, angular momentum is extracted from the BH, showing superradiance is still occurring, but it is being subdominant in terms of energy exchanges between the BH and the bulk AdS spacetime.
\end{itemize}

A feature that remains intriguing is that we have to consider RBCs, other than Dirichlet, to find superradiant modes in the BTZ BH. This seems to be a specific feature of this (2+1)-dimensional spacetime, since superradiance may be found in other asymptotically AdS BHs, such as the Kerr-AdS BH with Dirichlet boundary conditions \cite{Cardoso:2004hs}. This may be related to the fact that the horizon null generator $\chi = \partial_t + \Omega_{\mathcal{H}} \partial_{\varphi}$ is always timelike outside the horizon, in BTZ, but not for Kerr-AdS, say. However, the existence of the non-superradiant, unstable mode for certain RBCs is a generic consequence of the imposition of these boundary conditions and they should exist for other asymptotically AdS BHs.\footnote{We would like to thank J. E. Santos for drawing our attention to the fact that in AdS/CFT studies these RBCs have been considered, $e.g.$ in~\cite{Faulkner:2010fh}, as well as the instabilities they may generate.}

% ACKNOWLEDGEMENTS

\begin{acknowledgments}
We would like to thank F.~Finster, J.~Louko and T.~Sotiriou for useful discussions.  
The work of C.~D.\ was supported by the University of Pavia. 
The work of H.~F.\ was supported by the INFN postdoctoral fellowship ``Geometrical Methods in Quantum Field Theories and Applications'' and by the ``Progetto Giovani GNFM 2017 -- Wave propagation on lorentzian manifolds with boundaries and applications to algebraic QFT'' fostered by the National Group of Mathematical Physics (GNFM-INdAM).
C.H. is grateful to the INFN -- Sezione di Pavia for the kind hospitality during the realization of part of this work. C.~H. acknowledges funding from the FCT-IF programme. This work was partially supported by the H2020-MSCA-RISE-2015 Grant No. StronGrHEP-690904, and by the CIDMA project UID/MAT/04106/2013. We would like to acknowledge networking support by the COST Action GWverse
CA16104.
\end{acknowledgments}

% APPENDIX

%%%%%%%%%%%%%%%%%%%%%%%%%%%%%%
%%%%%%%%%%%%%%%%%%%%%%%%%%%%%%
%%%%%%%%%%%%%%%%%%%%%%%%%%%%%%
\appendix
%%%%%%%%%%%%%%%%%%%%%%%%%%%%%%
%%%%%%%%%%%%%%%%%%%%%%%%%%%%%%
%%%%%%%%%%%%%%%%%%%%%%%%%%%%%%

%%%%%%%%%%%%%%%%%%%%%%%%%%%%%%
%%%%%%%%%%%%%%%%%%%%%%%%%%%%%%
\section{Unstable modes in the Poincar\'e patch of anti-de Sitter}
\label{sec:AdS}
%%%%%%%%%%%%%%%%%%%%%%%%%%%%%%
%%%%%%%%%%%%%%%%%%%%%%%%%%%%%%

%\hf{This is a summary of the results of the AdS case. More detail can be added/removed later and its position on the paper is to be confirmed.}

The classical theory of a massive scalar field propagating in the Poincar\'{e} patch of AdS${}_{d+1}$ was analysed in Ref.~\cite{Dappiaggi:2016fwc}. Here, we reproduce the main results for 2+1 dimensions and discuss the existence of an unstable mode for a subclass of RBCs imposed at infinity.

The metric of the Poincar\'{e} patch of AdS${}_3$ is
\begin{equation}
\dd s^2 = \frac{\ell^2}{z^2} \left(- \dd t^2 + \dd z^2 + \dd x^2 \right) \, ,
\end{equation}
with $z>0$. The surface $z=0$ corresponds to part of the conformal boundary of AdS${}_3$ and these coordinates cover half of global AdS${}_3$.

We consider a massive scalar field $\Phi$ satisfying the Klein-Gordon equation,
\begin{equation}
\left(\Box - \frac{\mu^2}{\ell^2} \right)\Phi = 0 \, ,
\end{equation}
where $\mu^2/\ell^2$, with $\mu^2 > -1$, is the effective squared mass of the scalar field which, as with the BTZ spacetime, may include a coupling to the curvature, as the Ricci scalar $R=-6/\ell^2$ is a constant.

Using the ansatz $\Phi(t,z,x) = e^{-i\omega t + i k x} \phi(z)$, we obtain two linearly independent solutions for $\phi(z)$,
\begin{align}
\phi^{\rm (D)}(z) &= \sqrt{\frac{\pi}{2}} q^{-\nu} z J_{\nu}(qz) \, , \\
\phi^{\rm (N)}(z) &= \sqrt{\frac{\pi}{2}} q^{\nu} z J_{-\nu}(qz) \, ,
\end{align}
with $q^2 \equiv \omega^2 - k^2$. It is shown in \cite{Dappiaggi:2016fwc} that RBCs may be applied at $z=0$ only if $\nu \in (0,1)$, where $\nu \equiv \sqrt{1 + \mu^2}$.
In this case, the solution may be written as
\begin{equation}
\phi(z) = \cos(\zeta) \phi^{\rm (D)}(z) + \sin(\zeta) \phi^{\rm (N)}(z) \, ,
\end{equation}
with $\zeta \in [0,\pi)$ parametrizing the RBC. These solutions are square integrable eigensolutions of the eigenvalue problem associated with the field equation in $z$ and spectral parameter $q^2 > 0$. However, it can be shown that there is an extra, negative eigenvalue $q^2 = - \left[\cot(\pi-\zeta)\right]^{1/\nu}$ when $\zeta \in (\frac{\pi}{2},\pi)$ with square integrable eigensolution of the form $z K_{\nu}([\cot(\pi-\zeta)]^{1/(2\nu)} z)$. This is interpreted as a bound state mode solution, as it is exponentially suppressed as $z \to \infty$. 

\begin{figure}[t!]
	\centering
	\includegraphics[width=0.85\linewidth]{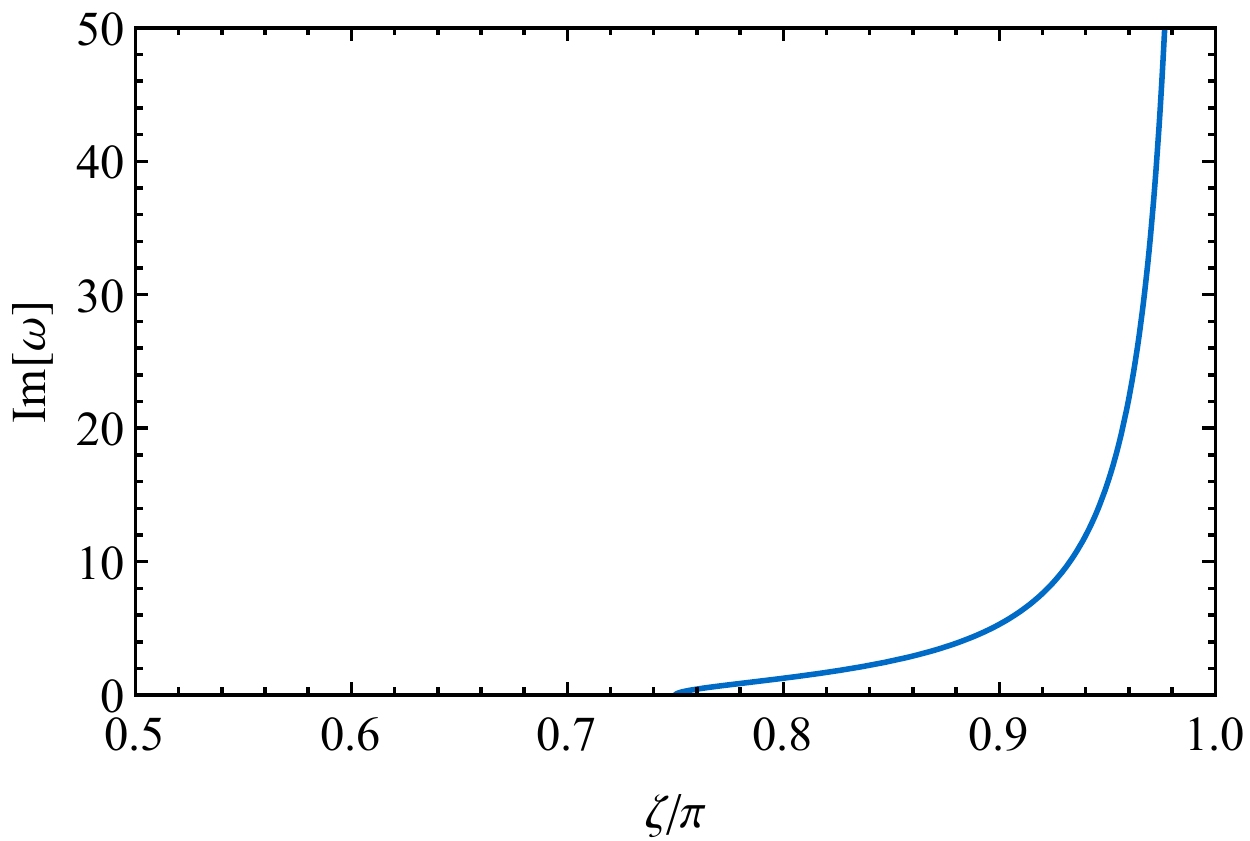}
	\caption{\label{fig:plot-AdS-Imomega}Imaginary part of the bound state frequency as a function of $\zeta/\pi$ for the Poincar\'e patch of AdS${}_3$ and scalar field with $\ell=1$, $\nu=1/3$ and $k=1$.}
\end{figure}

A given bound state mode solution with fixed $k$ and for fixed RBC has two possible frequencies,
\begin{equation} \label{eq:freqAdS}
\omega = \pm \sqrt{k^2 - \left[\cot(\pi-\zeta)\right]^{1/\nu}} \, .
\end{equation}
When $\zeta$ is such that $k^2 > \left[\cot(\pi-\zeta)\right]^{1/\nu}$, $\omega$ is real-valued and we can always take it to be positive without loss of generality. Otherwise, when $\zeta$ is such that $k^2 < \left[\cot(\pi-\zeta)\right]^{1/\nu}$, $\omega$ is purely imaginary and can have positive or negative imaginary part. In Fig.~\ref{fig:plot-AdS-Imomega}, as an example, we plot the imaginary part of this frequency for the bound state mode solution with positive imaginary part. 

We conclude that there exists a subclass of RBCs for which there is a mode solution which has a purely imaginary frequency with positive imaginary part, meaning that this mode grows exponentially in time. For these boundary conditions, of which Dirichlet ($\zeta = 0$) is not an example, the Poincar\'e patch of AdS${}_3$ is therefore linearly unstable to scalar mode perturbations. Note that this linear instability is conceptually different from the well-known weakly turbulent instability of AdS, which is nonlinear in nature and occurs even with Dirichlet boundary conditions (for a recent review see \cite{Martinon:2017ppj}).

Another feature of this unstable mode is that its frequency has imaginary part which diverges as $\zeta \to \pi^-$, as can be deduced from \eqref{eq:freqAdS}. It implies that the time scales involved in this instability are smaller for RBCs with values of $\zeta$ closer to $\pi$. As it is seen in the main text, this behavior also occurs for the BTZ BH and is a consequence of the AdS asymptotics of the spacetime.

% BIBLIOGRAPHY


\begin{thebibliography}{99}

%\cite{Brito:2015oca}
\bibitem{Brito:2015oca} 
  R.~Brito, V.~Cardoso and P.~Pani,
  %``Superradiance : Energy Extraction, Black-Hole Bombs and Implications for Astrophysics and Particle Physics,''
  Lect.\ Notes Phys.\  {\bf 906}, pp.1 (2015)
  %doi:10.1007/978-3-319-19000-6
  [gr-qc/1501.06570].
  
%\cite{Finster:2007fd}
\bibitem{Finster:2007fd} 
  F.~Finster, N.~Kamran, J.~Smoller and S.~T.~Yau,
  %``A Rigorous treatment of energy extraction from a rotating black hole,''
  Commun.\ Math.\ Phys.\  {\bf 287}, 829 (2009)
  [gr-qc/0701018].
  
  %\cite{Damour:1976kh}
\bibitem{Damour:1976kh}
  T.~Damour, N.~Deruelle and R.~Ruffini,
  %``On Quantum Resonances in Stationary Geometries,''
  Lett.\ Nuovo Cim.\  {\bf 15} (1976) 257.
  %doi:10.1007/BF02725534
  %%CITATION = doi:10.1007/BF02725534;%%
  %158 citations counted in INSPIRE as of 15 Oct 2017
  
  %\cite{Zouros:1979iw}
\bibitem{Zouros:1979iw} 
T.~J.~M.~Zouros and D.~M.~Eardley,
%``Instabilities Of Massive Scalar Perturbations Of A Rotating Black Hole,''
Annals Phys.\  {\bf 118}, 139 (1979).

%\cite{Detweiler:1980uk}
\bibitem{Detweiler:1980uk}
  S.~L.~Detweiler,
  %``Klein-gordon Equation And Rotating Black Holes,''
  Phys.\ Rev.\ D {\bf 22} (1980) 2323.
  %doi:10.1103/PhysRevD.22.2323
  %%CITATION = doi:10.1103/PhysRevD.22.2323;%%
  %221 citations counted in INSPIRE as of 15 Oct 2017

%\cite{Furuhashi:2004jk}
\bibitem{Furuhashi:2004jk}
  H.~Furuhashi and Y.~Nambu,
  %``Instability of massive scalar fields in Kerr-Newman space-time,''
  Prog.\ Theor.\ Phys.\  {\bf 112} (2004) 983
  %doi:10.1143/PTP.112.983
  [gr-qc/0402037].
  %%CITATION = doi:10.1143/PTP.112.983;%%
  %116 citations counted in INSPIRE as of 21 Oct 2017
	
%\cite{Dolan:2007mj}
\bibitem{Dolan:2007mj} 
S.~R.~Dolan,
%``Instability of the massive Klein-Gordon field on the Kerr spacetime,''
Phys.\ Rev.\ D {\bf 76}, 084001 (2007)
[arXiv:0705.2880 [gr-qc]].

%\cite{Rosa:2009ei}
\bibitem{Rosa:2009ei}
  J.~G.~Rosa,
  %``The Extremal black hole bomb,''
  JHEP {\bf 1006} (2010) 015
  %doi:10.1007/JHEP06(2010)015
  [arXiv:0912.1780 [hep-th]].
  %%CITATION = doi:10.1007/JHEP06(2010)015;%%
  %60 citations counted in INSPIRE as of 21 Oct 2017

%\cite{Pani:2012vp}
\bibitem{Pani:2012vp}
  P.~Pani, V.~Cardoso, L.~Gualtieri, E.~Berti and A.~Ishibashi,
  %``Black hole bombs and photon mass bounds,''
  Phys.\ Rev.\ Lett.\  {\bf 109} (2012) 131102
 % doi:10.1103/PhysRevLett.109.131102
  [arXiv:1209.0465 [gr-qc]].
  %%CITATION = doi:10.1103/PhysRevLett.109.131102;%%
  %101 citations counted in INSPIRE as of 19 Oct 2017


%\cite{Dolan:2012yt}
\bibitem{Dolan:2012yt}
  S.~R.~Dolan,
  %``Superradiant instabilities of rotating black holes in the time domain,''
  Phys.\ Rev.\ D {\bf 87} (2013) no.12,  124026
  %doi:10.1103/PhysRevD.87.124026
  [arXiv:1212.1477 [gr-qc]].
  %%CITATION = doi:10.1103/PhysRevD.87.124026;%%
  %95 citations counted in INSPIRE as of 19 Oct 2017

%\cite{Witek:2012tr}
\bibitem{Witek:2012tr}
  H.~Witek, V.~Cardoso, A.~Ishibashi and U.~Sperhake,
  %``Superradiant instabilities in astrophysical systems,''
  Phys.\ Rev.\ D {\bf 87} (2013) no.4,  043513
  %doi:10.1103/PhysRevD.87.043513
  [arXiv:1212.0551 [gr-qc]].
  %%CITATION = doi:10.1103/PhysRevD.87.043513;%%
  %97 citations counted in INSPIRE as of 19 Oct 2017

%\cite{Brito:2014wla}
\bibitem{Brito:2014wla}
  R.~Brito, V.~Cardoso and P.~Pani,
  %``Black holes as particle detectors: evolution of superradiant instabilities,''
  Class.\ Quant.\ Grav.\  {\bf 32} (2015) no.13,  134001
  %doi:10.1088/0264-9381/32/13/134001
  [arXiv:1411.0686 [gr-qc]].
  %%CITATION = doi:10.1088/0264-9381/32/13/134001;%%
  %36 citations counted in INSPIRE as of 19 Oct 2017
  
  %\cite{Hod:2016iri}
\bibitem{Hod:2016iri}
  S.~Hod,
  %``The superradiant instability regime of the spinning Kerr black hole,''
  Phys.\ Lett.\ B {\bf 758} (2016) 181
  %doi:10.1016/j.physletb.2016.05.012
  [arXiv:1606.02306 [gr-qc]].
  %%CITATION = doi:10.1016/j.physletb.2016.05.012;%%
  %16 citations counted in INSPIRE as of 21 Oct 2017

%\cite{Huang:2016qnk}
\bibitem{Huang:2016qnk}
  Y.~Huang and D.~J.~Liu,
  %``Scalar clouds and the superradiant instability regime of Kerr-Newman black hole,''
  Phys.\ Rev.\ D {\bf 94} (2016) no.6,  064030
  %doi:10.1103/PhysRevD.94.064030
  [arXiv:1606.08913 [gr-qc]].
  %%CITATION = doi:10.1103/PhysRevD.94.064030;%%
  %8 citations counted in INSPIRE as of 21 Oct 2017

  
  %\cite{East:2017mrj}
\bibitem{East:2017mrj}
  W.~E.~East,
  %``Superradiant instability of massive vector fields around spinning black holes in the relativistic regime,''
  Phys.\ Rev.\ D {\bf 96} (2017) no.2,  024004
  %doi:10.1103/PhysRevD.96.024004
  [arXiv:1705.01544 [gr-qc]].
  %%CITATION = doi:10.1103/PhysRevD.96.024004;%%
  %6 citations counted in INSPIRE as of 19 Oct 2017


%\cite{East:2017ovw}
\bibitem{East:2017ovw}
  W.~E.~East and F.~Pretorius,
  %``Superradiant Instability and Backreaction of Massive Vector Fields around Kerr Black Holes,''
  Phys.\ Rev.\ Lett.\  {\bf 119} (2017) no.4,  041101
  %doi:10.1103/PhysRevLett.119.041101
  [arXiv:1704.04791 [gr-qc]].
  %%CITATION = doi:10.1103/PhysRevLett.119.041101;%%
  %14 citations counted in INSPIRE as of 15 Oct 2017

%\cite{Herdeiro:2017phl}
\bibitem{Herdeiro:2017phl}
  C.~A.~R.~Herdeiro and E.~Radu,
  %``Kerr black holes with synchronised hair: an analytic model and dynamical formation,''
  arXiv:1706.06597 [gr-qc].
  %%CITATION = ARXIV:1706.06597;%%
  %2 citations counted in INSPIRE as of 15 Oct 2017

  
  
  
  %\cite{Herdeiro:2014goa}
\bibitem{Herdeiro:2014goa} 
  C.~A.~R.~Herdeiro and E.~Radu,
  %``Kerr black holes with scalar hair,''
  Phys.\ Rev.\ Lett.\  {\bf 112}, 221101 (2014)
  %doi:10.1103/PhysRevLett.112.221101
  [gr-qc/1403.2757].

%\cite{Herdeiro:2015gia}
\bibitem{Herdeiro:2015gia} 
  C.~Herdeiro and E.~Radu,
  %``Construction and physical properties of Kerr black holes with scalar hair,''
  Class.\ Quant.\ Grav.\  {\bf 32}, 14, 144001 (2015)
  %doi:10.1088/0264-9381/32/14/144001
  [gr-qc/1501.04319].
  
  %\cite{Herdeiro:2016tmi}
\bibitem{Herdeiro:2016tmi} 
  C.~Herdeiro, E.~Radu and H.~Runarsson,
  %``Kerr black holes with Proca hair,''
  Class.\ Quant.\ Grav.\  {\bf 33}, no. 15, 154001 (2016)
  [arXiv:1603.02687 [gr-qc]].

%\cite{Cardoso:2004hs}
\bibitem{Cardoso:2004hs}
  V.~Cardoso and O.~J.~C.~Dias,
  %``Small Kerr-anti-de Sitter black holes are unstable,''
  Phys.\ Rev.\ D {\bf 70} (2004) 084011
  %doi:10.1103/PhysRevD.70.084011
  [hep-th/0405006].
  %%CITATION = doi:10.1103/PhysRevD.70.084011;%%
  %151 citations counted in INSPIRE as of 19 Oct 2017
  
  %\cite{Li:2012rx}
\bibitem{Li:2012rx}
  R.~Li,
  %``Superradiant instability of charged massive scalar field in Kerr-Newman-anti-de Sitter black hole,''
  Phys.\ Lett.\ B {\bf 714} (2012) 337
  %doi:10.1016/j.physletb.2012.07.015
  [arXiv:1205.3929 [gr-qc]].
  %%CITATION = doi:10.1016/j.physletb.2012.07.015;%%
  %16 citations counted in INSPIRE as of 21 Oct 2017
  
  %\cite{Cardoso:2013pza}
\bibitem{Cardoso:2013pza}
  V.~Cardoso, î.~J.~C.~Dias, G.~S.~Hartnett, L.~Lehner and J.~E.~Santos,
  %``Holographic thermalization, quasinormal modes and superradiance in Kerr-AdS,''
  JHEP {\bf 1404} (2014) 183
  %doi:10.1007/JHEP04(2014)183
  [arXiv:1312.5323 [hep-th]].
  %%CITATION = doi:10.1007/JHEP04(2014)183;%%
  %55 citations counted in INSPIRE as of 21 Oct 2017
  
  %\cite{Wang:2015fgp}
\bibitem{Wang:2015fgp} 
  M.~Wang and C.~Herdeiro,
  %``Maxwell perturbations on KerrÐantiÐde Sitter black holes: Quasinormal modes, superradiant instabilities, and vector clouds,''
  Phys.\ Rev.\ D {\bf 93}, 6, 064066 (2016)
  %doi:10.1103/PhysRevD.93.064066
  [gr-qc/1512.02262].
 

%\cite{Delice:2015zga}
\bibitem{Delice:2015zga}
  ….~Delice and T.~Dur?ut,
  %``Superradiance Instability of Small Rotating AdS Black Holes in Arbitrary Dimensions,''
  Phys.\ Rev.\ D {\bf 92} (2015) no.2,  024053
  %doi:10.1103/PhysRevD.92.024053
  [arXiv:1503.05818 [gr-qc]].
  %%CITATION = doi:10.1103/PhysRevD.92.024053;%%
  %5 citations counted in INSPIRE as of 21 Oct 2017


%\cite{Banados:1992wn}
\bibitem{Banados:1992wn} 
  M.~Banados, C.~Teitelboim and J.~Zanelli,
  %``The Black hole in three-dimensional space-time,''
  Phys.\ Rev.\ Lett.\  {\bf 69}, 1849 (1992)
  %doi:10.1103/PhysRevLett.69.1849
  [hep-th/9204099].

%\cite{Banados:1992gq}
\bibitem{Banados:1992gq} 
  M.~Banados, M.~Henneaux, C.~Teitelboim and J.~Zanelli,
  %``Geometry of the (2+1) black hole,''
  Phys.\ Rev.\ D {\bf 48}, 1506 (1993)
  Erratum: [Phys.\ Rev.\ D {\bf 88}, 069902 (2013)]
  %doi:10.1103/PhysRevD.48.1506, 10.1103/PhysRevD.88.069902
  [gr-qc/9302012].
  
    %\cite{Ortiz:2011wd}
\bibitem{Ortiz:2011wd} 
  L.~Ortiz,
  %``No superradiance for the scalar field in the BTZ black hole with reflexive boundary conditions,''
  Phys.\ Rev.\ D {\bf 86}, 047703 (2012)
  %doi:10.1103/PhysRevD.86.047703
  [hep-th/1110.2555].
  
  %\cite{Wang:2015goa}
\bibitem{Wang:2015goa} 
  M.~Wang, C.~Herdeiro and M.~O.~P.~Sampaio,
  %``Maxwell perturbations on asymptotically antiÐde Sitter spacetimes: Generic boundary conditions and a new branch of quasinormal modes,''
  Phys.\ Rev.\ D {\bf 92}, 12, 124006 (2015)
  %doi:10.1103/PhysRevD.92.124006
  [gr-qc/1510.04713].

%\cite{Dappiaggi:2016fwc}
\bibitem{Dappiaggi:2016fwc} 
  C.~Dappiaggi and H.~R.~C.~Ferreira,
  %``Hadamard states for a scalar field in anti–de Sitter spacetime with arbitrary boundary conditions,''
  Phys.\ Rev.\ D {\bf 94}, no. 12, 125016 (2016)
  [arXiv:1610.01049 [gr-qc]].

%\cite{Ferreira:2017cta}
\bibitem{Ferreira:2017cta} 
  H.~R.~C.~Ferreira and C.~A.~R.~Herdeiro,
  %``Stationary scalar clouds around a BTZ black hole,''
  Phys.\ Lett.\ B {\bf 773}, 129 (2017)
  [arXiv:1707.08133 [gr-qc]].
  
  %\cite{Iizuka:2015vsa}
\bibitem{Iizuka:2015vsa}
  N.~Iizuka, A.~Ishibashi and K.~Maeda,
  %``A rotating hairy AdS$_{3}$ black hole with the metric having only one Killing vector field,''
  JHEP {\bf 1508} (2015) 112
  %doi:10.1007/JHEP08(2015)112
  [arXiv:1505.00394 [hep-th]].
  %%CITATION = doi:10.1007/JHEP08(2015)112;%%
  %4 citations counted in INSPIRE as of 19 Oct 2017

%\cite{Bussola:2017wki}
\bibitem{Bussola:2017wki} 
  F.~Bussola, C.~Dappiaggi, H.~R.~C.~Ferreira and I.~Khavkine,
  %``Ground state for a massive scalar field in BTZ spacetime with Robin boundary conditions,''
  arXiv:1708.00271 [gr-qc].

%\cite{Dias:2011at}
\bibitem{Dias:2011at}
  O.~J.~C.~Dias, G.~T.~Horowitz and J.~E.~Santos,
  %``Black holes with only one Killing field,''
  JHEP {\bf 1107} (2011) 115
  %doi:10.1007/JHEP07(2011)115
  [arXiv:1105.4167 [hep-th]].
  %%CITATION = doi:10.1007/JHEP07(2011)115;%%
  %97 citations counted in INSPIRE as of 21 Oct 2017
  
  %\cite{Hod:2012px}
\bibitem{Hod:2012px} 
  S.~Hod,
  %``Stationary Scalar Clouds Around Rotating Black Holes,''
  Phys.\ Rev.\ D {\bf 86}, 104026 (2012)
  Erratum: [Phys.\ Rev.\ D {\bf 86}, 129902 (2012)]
  %doi:10.1103/PhysRevD.86.129902, 10.1103/PhysRevD.86.104026
  [gr-qc/1211.3202].

  
  %\cite{Hod:2013zza}
\bibitem{Hod:2013zza} 
  S.~Hod,
  %``Stationary resonances of rapidly-rotating Kerr black holes,''
  Eur.\ Phys.\ J.\ C {\bf 73}, 4, 2378 (2013)
  %doi:10.1140/epjc/s10052-013-2378-x
  [gr-qc/1311.5298].
  
  %\cite{Hod:2014baa}
\bibitem{Hod:2014baa}
  S.~Hod,
  %``Kerr-Newman black holes with stationary charged scalar clouds,''
  Phys.\ Rev.\ D {\bf 90} (2014) no.2,  024051
 % doi:10.1103/PhysRevD.90.024051
  [arXiv:1406.1179 [gr-qc]].
  %%CITATION = doi:10.1103/PhysRevD.90.024051;%%
  %45 citations counted in INSPIRE as of 19 Oct 2017
  
  %\cite{Benone:2014ssa}
\bibitem{Benone:2014ssa}
  C.~L.~Benone, L.~C.~B.~Crispino, C.~Herdeiro and E.~Radu,
  %``Kerr-Newman scalar clouds,''
  Phys.\ Rev.\ D {\bf 90} (2014) no.10,  104024
  %doi:10.1103/PhysRevD.90.104024
  [arXiv:1409.1593 [gr-qc]].
  %%CITATION = doi:10.1103/PhysRevD.90.104024;%%
  %58 citations counted in INSPIRE as of 19 Oct 2017


%\cite{Wilson-Gerow:2015esa}
\bibitem{Wilson-Gerow:2015esa}
  J.~Wilson-Gerow and A.~Ritz,
  %``Black hole energy extraction via a stationary scalar analog of the Blandford-Znajek mechanism,''
  Phys.\ Rev.\ D {\bf 93} (2016) no.4,  044043
  %doi:10.1103/PhysRevD.93.044043
  [arXiv:1509.06681 [hep-th]].
  %%CITATION = doi:10.1103/PhysRevD.93.044043;%%
  %7 citations counted in INSPIRE as of 19 Oct 2017

%\cite{Bernard:2016wqo}
\bibitem{Bernard:2016wqo}
  C.~Bernard,
  %``Stationary charged scalar clouds around black holes in string theory,''
  Phys.\ Rev.\ D {\bf 94} (2016) no.8,  085007
  %doi:10.1103/PhysRevD.94.085007
  [arXiv:1608.05974 [gr-qc]].
  %%CITATION = doi:10.1103/PhysRevD.94.085007;%%
  %3 citations counted in INSPIRE as of 19 Oct 2017

%\cite{Sakalli:2016xoa}
\bibitem{Sakalli:2016xoa}
  I.~Sakalli and G.~Tokgoz,
  %``Stationary Scalar Clouds Around Maximally Rotating Linear Dilaton Black Holes,''
  Class.\ Quant.\ Grav.\  {\bf 34} (2017) no.12,  125007
 % doi:10.1088/1361-6382/aa6858
  [arXiv:1610.09329 [gr-qc]].
  %%CITATION = doi:10.1088/1361-6382/aa6858;%%

%\cite{Hod:2016lgi}
\bibitem{Hod:2016lgi}
  S.~Hod,
  %``Spinning Kerr black holes with stationary massive scalar clouds: The large-coupling regime,''
  JHEP {\bf 2017} (2017) 030
  [arXiv:1612.00014 [hep-th]].


%\cite{Gibbons:2008zi}
\bibitem{Gibbons:2008zi}
  G.~W.~Gibbons, C.~A.~R.~Herdeiro, C.~M.~Warnick and M.~C.~Werner,
  %``Stationary Metrics and Optical Zermelo-Randers-Finsler Geometry,''
  Phys.\ Rev.\ D {\bf 79} (2009) 044022
  %doi:10.1103/PhysRevD.79.044022
  [arXiv:0811.2877 [gr-qc]].
  %%CITATION = doi:10.1103/PhysRevD.79.044022;%%
  %66 citations counted in INSPIRE as of 19 Oct 2017

%\cite{Breitenlohner:1982jf}
\bibitem{Breitenlohner:1982jf} 
P.~Breitenlohner and D.~Z.~Freedman,
%``Stability in Gauged Extended Supergravity,''
Annals Phys.\  {\bf 144}, 249 (1982).

%\cite{NIST}
\bibitem{NIST}
F.~Olver,
{\it NIST Handbook of Mathematical Functions,}
Cambridge University Press (2010).


%\cite{Birmingham:2001hc}
\bibitem{Birmingham:2001hc} 
D.~Birmingham,
%``Choptuik scaling and quasinormal modes in the AdS/CFT correspondence,''
Phys.\ Rev.\ D {\bf 64}, 064024 (2001)
[hep-th/0101194].


%\cite{Faulkner:2010fh}
\bibitem{Faulkner:2010fh}
  T.~Faulkner, G.~T.~Horowitz and M.~M.~Roberts,
  %``New stability results for Einstein scalar gravity,''
  Class.\ Quant.\ Grav.\  {\bf 27} (2010) 205007
  %doi:10.1088/0264-9381/27/20/205007
  [arXiv:1006.2387 [hep-th]].
  %%CITATION = doi:10.1088/0264-9381/27/20/205007;%%
  %42 citations counted in INSPIRE as of 19 Oct 2017
	

%\cite{Martinon:2017ppj}
\bibitem{Martinon:2017ppj} 
  G.~Martinon,
  %``The instability of anti-de Sitter space-time,''
  arXiv:1708.05600 [gr-qc].  
 


  
\end{thebibliography}
\end{document}